\documentclass[journal]{IEEEtran}
\usepackage[utf8]{inputenc} %unicode support
\usepackage{graphicx}
\usepackage{subfigure}
\usepackage{amsmath}
\usepackage{multirow}
\usepackage[varg]{newtxmath}
\usepackage{booktabs}
\usepackage{amsmath}

\usepackage[psamsfonts]{amssymb}
\usepackage{amsxtra}
\usepackage{threeparttable}
\usepackage{romannum}
\usepackage{stfloats}
\usepackage{cite}
\usepackage{color, xcolor}
\usepackage{threeparttable}
\usepackage{ulem}
\usepackage{makecell}
\usepackage{algorithmic}
\usepackage{algorithm}

\ifCLASSINFOpdf
\else
\fi
\begin{document}

\title{A Squeeze-and-Excitation and Transformer based Cross-task Model for Environmental Sound Recognition}

%%%%%%%%%%%%%%%%%%%%%%%%%%%%%%%%%%%%%%%%%%%%%%
%%                                          %%
%% Enter the authors here                   %%
%%                                          %%
%% Specify information, if available,       %%
%% in the form:                             %%
%%   <key>={<id1>,<id2>}                    %%
%%   <key>=                                 %%
%% Comment or delete the keys which are     %%
%% not used. Repeat \author command as much %%
%% as required.                             %%
%%                                          %%
%%%%%%%%%%%%%%%%%%%%%%%%%%%%%%%%%%%%%%%%%%%%%%

\author{Jisheng Bai,~\IEEEmembership{Student~Member,~IEEE,}
	Jianfeng Chen,~\IEEEmembership{Senior Member,~IEEE,}
	Mou Wang,~\IEEEmembership{Student~Member,~IEEE},
	Muhammad Saad Ayub,
	and Qingli Yan
	%and~Xiao-Lei Zhang,~\IEEEmembership{Senior Member,~IEEE}
	% <-this % stops a space
	\thanks{This work is supported by the National Natural Science Foundation of China under Grant No.62071383 and the Key research and development plan of Shaanxi Province (2021NY-036). \textit{(Corresponding author: Jianfeng Chen.)} }
    \thanks{J. Bai, J. Chen, M. Wang and M. S. Ayub are with the Joint Laboratory of Environmental Sound Sensing, School of Marine Science and Technology, Northwestern Polytechnical University, Xi’an, China, and also with the LianFeng Acoustic Technologies Co., Ltd. Xi'an, China. (e-mail:  baijs@mail.nwpu.edu.cn; chenjf@nwpu.edu.cn; wangmou21@mail.nwpu.edu.cn; msaadayub@mail.nwpu.edu.cn)}
    \thanks{Qingli Yan is with the School of Computer Science $\&$ Technology, Xi'an University of Posts $\&$ Telecommunications, Xi'an, China (e-mail: yql@xupt.edu.cn)
	}% <-this % stops a space  ; xiaolei.zhang@nwpu.edu.cn
	% <-this % stops a space
}

%%%%%%%%%%%%%%%%%%%%%%%%%%%%%%%%%%%%%%%%%%%%%%
%%                                          %%
%% Enter the authors' addresses here        %%
%%                                          %%
%% Repeat \address commands as much as      %%
%% required.                                %%
%%                                          %%
%%%%%%%%%%%%%%%%%%%%%%%%%%%%%%%%%%%%%%%%%%%%%%

%%%%%%%%%%%%%%%%%%%%%%%%%%%%%%%%%%%%%%%%%%%%%%
%%                                          %%
%% Enter short notes here                   %%
%%                                          %%
%% Short notes will be after addresses      %%
%% on first page.                           %%
%%                                          %%
%%%%%%%%%%%%%%%%%%%%%%%%%%%%%%%%%%%%%%%%%%%%%%
\maketitle
\pagenumbering{arabic}
%\end{fmbox}% comment this for two column layout

%%%%%%%%%%%%%%%%%%%%%%%%%%%%%%%%%%%%%%%%%%%%%%
%%                                          %%
%% The Abstract begins here                 %%
%%                                          %%
%% Please refer to the Instructions for     %%
%% authors on http://www.biomedcentral.com  %%
%% and include the section headings         %%
%% accordingly for your article type.       %%
%%                                          %%
%%%%%%%%%%%%%%%%%%%%%%%%%%%%%%%%%%%%%%%%%%%%%%

\begin{abstract} % abstract
Environmental sound recognition (ESR) is an emerging research topic in audio pattern recognition.
Many tasks are presented to resort to computational models for ESR in real-life applications.
However, current models are usually designed for individual tasks, and are not robust and applicable to other tasks.
Cross-task models, which promote unified knowledge modeling across various tasks, have not been thoroughly investigated.
In this paper, we propose a cross-task model for three different tasks of ESR: acoustic scene classification, urban sound tagging, and anomalous sound detection.
An architecture named SE-Trans is presented that uses attention mechanism-based Squeeze-and-Excitation and Transformer encoder modules to learn channel-wise relationship and temporal dependencies of the acoustic features.
FMix is employed as the data augmentation method that improves the performance of ESR.
Evaluations for the three tasks are conducted on the recent databases of DCASE challenges. 
The experimental results show that the proposed cross-task model achieves state-of-the-art performance on all tasks. 
Further analysis demonstrates that the proposed cross-task model can effectively utilize acoustic knowledge across different ESR tasks.
	
%\parttitle{First part title} %if any
%Text for this section.

%\parttitle{Second part title} %if any
%Text for this section.
\end{abstract}

%%%%%%%%%%%%%%%%%%%%%%%%%%%%%%%%%%%%%%%%%%%%%%
%%                                          %%
%% The keywords begin here                  %%
%%                                          %%
%% Put each keyword in separate \kwd{}.     %%
%%                                          %%
%%%%%%%%%%%%%%%%%%%%%%%%%%%%%%%%%%%%%%%%%%%%%%


% Self-Supervised Vision-Based Detection of the Active Speaker as Support for Socially Aware Language Acquisition
%Anomalous Behaviors Detection in Moving Crowds Based on a Weighted Convolutional Autoencoder-Long Short-Term Memory Network

% MSC classifications codes, if any
%\begin{keyword}[class=AMS]
%\kwd[Primary ]{}
%\kwd{}
%\kwd[; secondary ]{}
%\end{keyword}

\IEEEpeerreviewmaketitle

%%%%%%%%%%%%%%%%%%%%%%%%%%%%%%%%%%%%%%%%%%%%%%
%%                                          %%
%% The Main Body begins here                %%
%%                                          %%
%% Please refer to the instructions for     %%
%% authors on:                              %%
%% http://www.biomedcentral.com/info/authors%%
%% and include the section headings         %%
%% accordingly for your article type.       %%
%%                                          %%
%% See the Results and Discussion section   %%
%% for details on how to create sub-sections%%
%%                                          %%
%% use \cite{...} to cite references        %%
%%  \cite{koon} and                         %%
%%  \cite{oreg,khar,zvai,xjon,schn,pond}    %%
%%  \nocite{smith,marg,hunn,advi,koha,mouse}%%
%%                                          %%
%%%%%%%%%%%%%%%%%%%%%%%%%%%%%%%%%%%%%%%%%%%%%%

%%%%%%%%%%%%%%%%%%%%%%%%% start of article main body
% <put your article body there>

%%%%%%%%%%%%%%%%
%% Background %%
%%
%\section*{Content}
%Text and results for this section, as per the individual journal's instructions for authors. %\cite{koon,oreg,khar,zvai,xjon,schn,pond,smith,marg,hunn,advi,koha,mouse}

\section{Introduction}
\label{sec:into}
%1. ESR 介绍、应用、意义
%2. DCASE促进该领域发展。DCASE包含ASC UST ASD。
% 都属于音频信号处理，虽然不同，有相似性。
%特征logmel，模型除了早期的传统机器学习方法，现在深度学习主导。CNN，CRNN，Transformer。数据增强mixup 作为主要模块
%3.不同领域都有研究通用性系统的工作，bert VIT-backbone PANN。 考虑通用性，和DCASE任务相似性，cross-task有意义。
%kong16-19的工作。
% 4。 Attention 各领域提升。
% Attention  (1 自上而下 top-down focus of attention MHSA 
% (2 自下而上  bottom-up Saliency-Based Attention task-dependent Winner-Take-All or Gating  SE
% 不同任务两种分别有用过。还没有在cross-task里研究attention
%5。 our cross-task for ASC UST ASD.  两种attention被探索。
%  自下而上 SE-特征显著性，
%自上而下MHSA-探索长序列以来建模。mixup——fmix
%6.contribution：
%cross-task/两种attention在不同任务上探索/fmix

% 修改图1
% 修改图2
% 修改图3
% 所有图表标题、描述和大小
% 所有数字加逗号 
% 图6 7 8重画
% 图9 图例
%数据集规范描述 development set training and testing sets √
% baseline systems 总分描述 √
% sec 5 总描述 cross-task subtask总描述 √

Humans can automatically recognize sounds, but it is challenging for machines.  
Audio pattern recognition (APR) is a growing research area where signal processing and machine learning methods are used to understand the surrounding sound.
APR is of great importance in automatic speech recognition, automatic music transcription, and environmental sound recognition (ESR). 
More recently, ESR has attracted much attention due to various applications are emerging in our daily life.
For example, heart sound has been used as a biometric to identify a person in a real-time authentication system \cite{PHUA2008906}. 
In surveillance systems, the detection of gunshots or glass breaking can be used to report danger in time \cite{7096526}. 
More meaningful applications are using speech and non-speech audio to detect COVID-19 \cite{DESHPANDE2022108289}, recognising crying and speech sounds of babies to protect and understand babies \cite{torres2017baby, asada2016modeling}, and monitoring depression with acoustic and visual features \cite{jan2017artificial}.

The Detection and Classification of Acoustic Scenes and Events (DCASE) challenges, which focus on machine listening research for acoustic environments, have dramatically promoted the development of ESR in the past few years \cite{7100934}.
The series of challenges on DCASE have provided a set of tasks, such as acoustic scene classification (ASC) \cite{7078982}, urban sound tagging (UST) \cite{cartwright2020sonyc, bai2022multimodal}, and anomalous sound detection (ASD) \cite{Kawaguchi_arXiv2021_01}, encouraging the participants to develop strong computational models \cite{8123864}.
The computational models for ESR usually consists of two stages: feature extraction, in which audio is transformed into a feature representation, and sound recognition, in which a mapping between the feature representations and labels of sound classes is learned by a classifier  \cite{virtanen2018computational}.
Log Mel spectrograms, which represent the audio signal using perceptually motivated frequency scales\cite{9524590}, have been the predominant feature representations in this field \cite{BAELDE201982, ALAMIR2021107829}.
Recently, deep learning methods have achieved great success in many pattern recognition fields, including image classification , speech processing, natural language processing (NLP), and ESR \cite{szegedy2016rethinking, luo2019conv, sutskever2014sequence, 8667664, 9747306}.
Deep neural networks (DNNs) based classifiers, such as convolutional neural networks (CNNs), convolutional recurrent neural networks (CRNNs), and Transformer, have become dominant approaches in ESR and outperform the conventional machine learning methods \cite{vuegen2013mfcc, uzkent2012non}. 
Moreover, data augmentation has been a necessary part of an ESR models to mitigate overfitting during the training stage, and further improves the performance 
\cite{mushtaq2021spectral}.

In DCASE challenges, ESR models are usually developed for concrete and individual tasks, 
but the unification of the models designed for these tasks has not been adequately studied.
Unified models across different domains will facilitate knowledge modeling from these fields, and such unification has been explored in NLP (e.g. Bert \cite{devlin2018bert}), computer vision (CV) (e.g. ViT backbone \cite{dosovitskiy2020image}), and APR (e.g. PANN \cite{kong2020panns}).
Considering the above reason, it is of great importance to explore a generic model that has robust performance and wide applicability for different ESR tasks, where the common acoustic knowledge can be modeled in the model.
The models across several tasks in ESR are called \emph{cross-task} models \cite{kong2019cross}.
They have investigated the performance of CNNs-based models for several tasks of DCASE challenges in 2018 and 2019, and found that a 9-layer CNN with average pooling is a good model for most of the tasks.

Recently, attention mechanism has attracted tremendous interest in numerous pattern recognition fields as a component of DNNs.
Implementing attention can get more powerful representations by allowing the networks to dynamically pay attention to the effective information of the signal.
For speech emotion recognition, CNN with self-attention mechanism has been used to obtain abundant
emotional information on frame-level \cite{jiang2021convolutional}.
In ESR, different attention mechanisms have been applied for environmental sound classification \cite{zhang2021attention}, sound event detection\cite{kong2020sound}, and ASC \cite{ren2018attention}.
Yet, to the best of our knowledge, attention mechanism has not been studied or exploited in cross-task models.

In this paper, we propose a cross-task model based on an attention mechanism for modeling on three subtasks of ESR, i.e., ASC, UST, and ASD.
First, we propose to use \emph{Squeeze-and-Excitation} (SE) modules after convolution layers in the main architecture \cite{hu2018squeeze}. 
The model is allowed to learn the importance of channel-wise features, and the acoustic information between channels is effectively enhanced.
Next, we adopt the Transformer encoder in our architecture, where the multi-head self-attention (MHSA) mechanism is applied to efficiently model and process the temporal dependencies.
In our proposed model, the main architecture is based on SE and Transformer encoder modules, which is called \emph{SE-Trans}.
Besides, we proposed to use FMix as the data augmentation method, which can effectively augment the training data by randomly mixing irregular areas of two samples.
The main contributions of this paper can be summarized as follows:
\begin{itemize}
	\item We propose a cross-task model that can generally model various audio patterns across several tasks of ESR.
	\item We incorporate two modules with attention mechanisms, i.e., SE and Transformer encoder, into the main architecture to enhance channel-wise acoustic information and catch temporal dependencies.
	\item We firstly introduce FMix into ESR as the data augmentation method and find FMix can effectively augment the training data and significantly improve the performance of ESR tasks.
	\item We conduct experiments on the latest dataset of DACSE challenges. And results show that our cross-task model achieves state-of-the-art performance for ASC, UST, and ASD.
\end{itemize}

The rest of the paper is organized as follows: 
Section \ref{sec:related works} reviews the related works of this paper. 
Section \ref{sec:method} describes the detailed parts of the proposed cross-task model.
Section \ref{sec:exp} presents the experiments, the results and discussions are given in Section \ref{sec:results}. 
Finally, we conclude this paper in Section \ref{sec:conclusion}.

\begin{figure*}[t!]
	\centering
	\includegraphics[width=130mm]{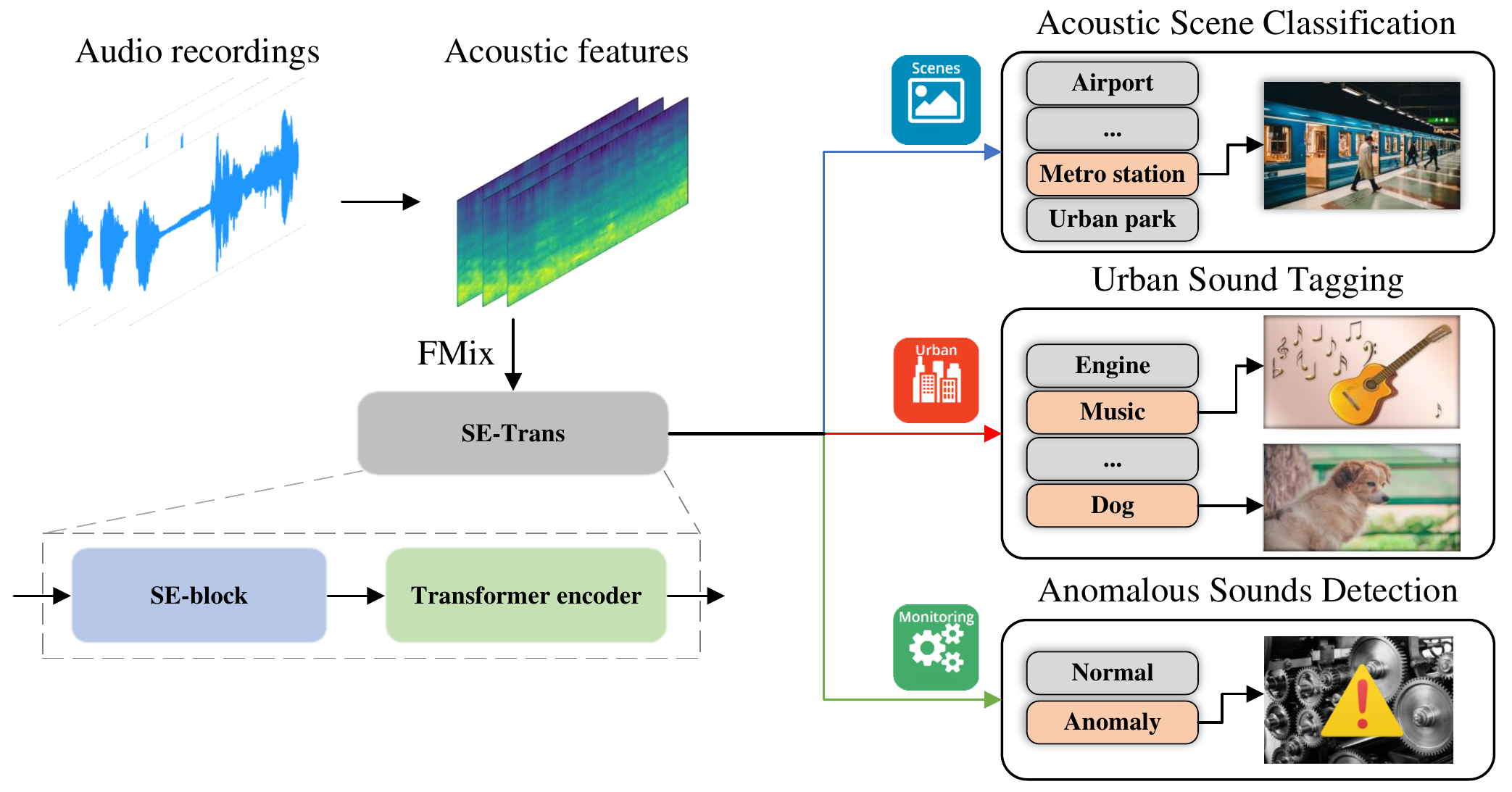}
	\caption{The processing stages of the proposed cross-task model.}
	\label{fig:System_overview}
\end{figure*}

\section{Task description and related works}
\label{sec:related works}
%1 ESR 重要领域 ASC UST ASD
%2. ASC introduction/(methods attention,problems?)
%3. UST introduction/(methods attention,problems?)
%4. ASD introduction/(methods attention,problems?)
%5. kong Cross-task introduction
%6. data augmentation
%增强 cutout,random erase --SpecAugment
%     MSDA     mixup cutmix——fmix

DCASE challenges have been successfully held seven times since 2013.
Several tasks closed to different aspects of real-life applications are brought by the organizers, providing public datasets, metrics, and evaluation frameworks \cite{8123864}.
Among the tasks, ASC, UST, and ASD, which are consecutively organized for at least 2 years, have attracted much interest.

ASC means to identify an acoustic scene among the predefined classes in the environment, using signal processing and machine learning methods.
The acoustic scenes are usually recorded in real-life environments such as squares, streets, and restaurants.
Many applications can potentially use ASC systems, e.g., wearable devices \cite{harma2003techniques}, robotics \cite{martinson2007robotic}, and smart home devices \cite{mesaros2018acoustic}.
Early works primarily focused on using conventional machine learning methods, such as Gaussian mixture models and support vector machines.
Gradually, CNNs-based approaches have become the mainstream for designing systems and achieved the top performance in recent years \cite{Chen2019, Suh2020}.

The goal of UST is to predict whether an urban sound exists in a recording.
Various urban sounds occur around us all the time when living in the big cities.
Some urban sounds can be harmful if we are exposed to them for a long time, such as traffic noise.
Sounds of New York City (SONYC) is a research project investigating data-driven approaches to mitigate urban noise pollution \cite{Bello2019sonyc}.
SONYC has collected over 100 million recordings between 2016 and 2019, and the researchers held the UST tasks in DCASE 2019 and DCASE 2020 to resort to computational methods for automatically monitoring noise pollution. 
Both of the champions in DCASE 2019 and DCASE 2020 UST tasks used CNNs as the primary architecture of the classifiers \cite{Adapa2019, Iqbal2020}.

ASD aims to detect anomalous acoustic signals in particular environments. 
ASD for machine condition monitoring (MCM) is an emerging task to identify whether the sound produced from a target machine is normal or anomalous.
Some of the successful methods in ASD can reduce the loss caused by machine damage and speed up the use of essential technologies in industry.
In factories, anomalous sounds rarely occur and are often unavailable. 
The main challenge is to detect unknown anomalous sounds under the condition that only normal sound samples are provided.
To address this problem, some researchers organized the tasks named "unsupervised detection of anomalous sounds for MCM" in DCASE 2020 and DCASE 2021 \cite{Koizumi_DCASE2020_01, Kawaguchi_arXiv2021_01}.
In \cite{dohi2021flow}, the authors proposed a self-supervised density estimation
method using normalizing flows and machine IDs to detect anomalies.
A MobileFaceNet was trained in a self-supervised learning manner to detect anomalous sound and achieved great performance in DCASE 2021 \cite{MoritaSECOM2021}.

Recently, using deep learning in the above tasks of ESR has become the trend and achieved state-of-the-art performance.
Yet, these methods mostly focus on specific tasks, we have observed little relevant literature on studying unification in this field.
The study of cross-task tries to find a general model (system) that can perform well for various tasks in ESR, and deep learning based cross-task models have been investigated.
Kong et al. proposed DNNs based baseline with the same structure for the DCASE 2016 challenge \cite{Kong2016}.
The DNNs take Mel-filter bank features as input and outperform the official baselines in many but not all tasks.
In DCASE 2018 challenge, they created a cross-task baseline system for all five tasks based on CNNs \cite{Kong2018}.
CNNs with 4 layers and 8 layers are investigated through performance for various tasks with the same configuration of neural networks. 
They found that deeper CNN with 8 layers performs better than CNN with 4 layers on almost all tasks.
Further, Kong et al. proposed generic cross-task baseline systems, where CNNs with 5, 9, and 13 layers are studied \cite{kong2019cross}.
The results of the CNN models on five tasks of DCASE 2019 showed that the 9-layer CNN with average pooling could achieve good performance in general.

CNNs are designed specifically for images and have become the dominant models in CV \cite{chaudhari2021attentive}. 
Because of its great ability to extract features from spectrograms, CNNs have also been the most popular architecture in ESR.
Most recently, CNNs have incorporated many attention mechanisms, such as SE and self-attention,
to focus on key parts, catch long-range dependencies of the input, and reduce computational complexity.
The combination of CNNs and attention has outperformed many of the previous methods and become a new research direction.
SENet \cite{hu2018squeeze} was proposed to explicitly learn the relationships between the channels and pay more attention to the more important feature maps.
Implementing SE can effectively improve the performance of classification with little increase of the parameters.
In addition, Transformer has extensively achieved state-of-the-art performance in many research domains. 
The MHSA modules in Transformer can model the input of long sequences and process in parallel.
A combination of CNNs and Transformers has been proposed to model both local and global dependencies of an audio sequence for speech recognition \cite{gulati2020conformer}.

Most state-of-the-art approaches in ESR use data augmentation in the training stage to overcome the overfitting problem caused by the lack of environmental sound.
These data augmentation methods can be categorized into two classes, depending on the representation format of sound.
In the first class, the methods operate on the sound waveform, including changing the speed, volume, and pitch etc \cite{mushtaq2021spectral}.
The drawback of these methods is that they need complex operations and a lot of time to generate enough samples.
Another class is usually carried out on the spectrogram without taking too much time to get enough data.
These methods are derived from the methods for augmenting images, using two main strategies.
One strategy does data augmentation by removing or masking some information on the images or spectrograms (e.g. cutout\cite{devries2017improved} or SpecAugment \cite{park2019specaugment}).
The above strategy is undesirable for some classification tasks because it may lose part of the information during training.
Another strategy called mixed sample data augmentation (MSDA) augments the training data by combining samples according to a specific policy, such as mixup\cite{zhang2017mixup} and FMix \cite{harris2020fmix}.
The MSDA methods can generate more unseen data, force the model to learn more robust features, and finally improve the performance of classification.

\section{Proposed Method}\label{sec:proposed method}
\label{sec:method}
\subsection{Model overview} 
The processing stages of the proposed cross-task model are shown in Fig. \ref{fig:System_overview}. 
First, the model takes audio recordings as input and transforms them into acoustic features.
Then, FMix is applied to the acoustic features to generate mixed acoustic features.
Next, the SE-Trans, which consists of SE-blocks and a Transformer encoder, is trained to recognize the acoustic features under different situations. 
For ASC, each audio recording will be recognized as a specific acoustic scene.
For UST, each audio recording will be tagged with different urban sound classes.
And for ASD, each audio recording will be annotated as normal or anomaly.

\subsection{SE-block}
The first part of the proposed model is SE-blocks.  
We denote $\mathbf{X} \in \mathbb{R}^{T\times F}$ as an acoustic feature transformed from an audio recording $x$, where $T$ is the number
of time frames and $F$ is the number of frequency bins.
Then $\mathbf{X}$ is further reshaped into $\mathbf{X} \in \mathbb{R}^{T\times F \times 1}$, which is consecutively processed by convolutional layers (Conv), batch normalization (BN), SE layers, and rectified linear unit (ReLU) for two times in each SE-block.

\begin{figure*}[htbp]
\centering
	\includegraphics[width=120mm]{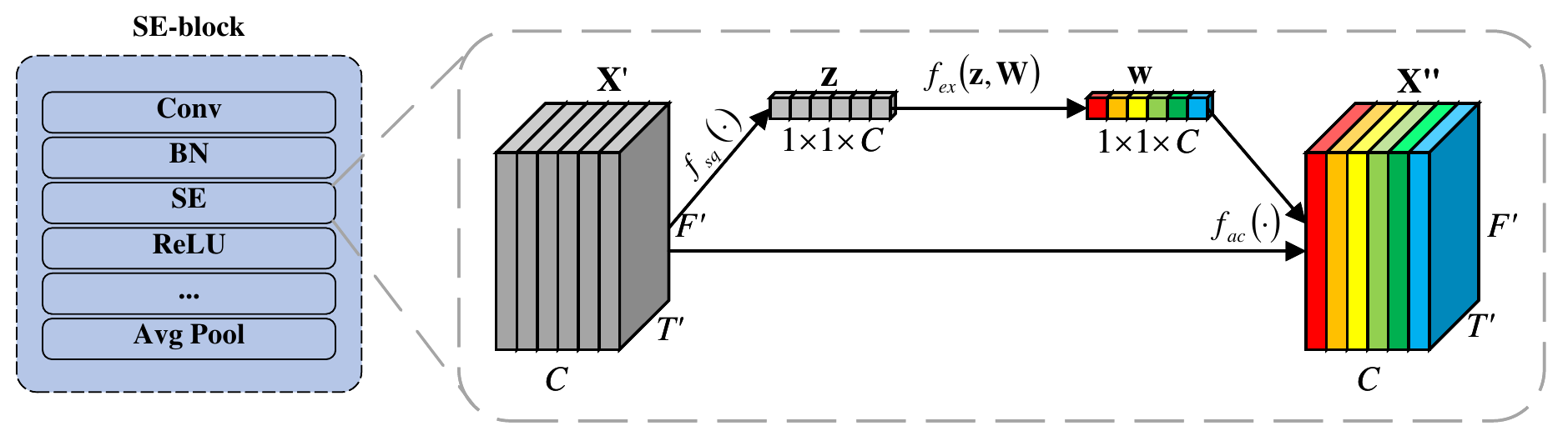}
	\caption{A flowchart of the SE layer in the SE-block. Convolutional layers (Conv), batch normalization (BN), rectified linear units (ReLU), average pooling layer (Avg Pool).}
	\label{fig:SE block}
\end{figure*}

We assume that the output of BN is $\mathbf{X'}=[\mathbf{x'}_{1},\mathbf{x'}_{2},\cdots,\mathbf{x'}_{c}]\in \mathbb{R}^{T'\times F' \times C}$, where $C$ is the number of channels of the convolutional layer,  $T'$ is the number of time frames, and $F'$ is the number of frequency bins. 
In an SE layer, $\mathbf{X'}$ is first \emph{squeezed} through time-frequency dimensions $T'\times F'$ in each channel:

\begin{equation}
z_{c}=f_{sq}\left(\mathbf{x'}_{c}\right)=\frac{1}{T' \times F'} \sum_{i=1}^{T'} \sum_{j=1}^{F'} \mathbf{x'}_{c}(i, j),
\end{equation}

where $f_{sq}$ is the \emph{global average pooling} function, and $z_{c}$ is the channel-wise value of squeezed vector $\mathbf{z}\in \mathbb{R}^{C}$.
A channel-wise relationship is then \emph{excited} from $\mathbf{z}$ by a gating mechanism: 
\begin{equation}
\mathbf{w}=f_{ex}(\mathbf{z}, \mathbf{W})=\sigma\left(\mathbf{W_{2}} \operatorname{ReLU}\left(\mathbf{W_{1}}\mathbf{z}\right)\right),
\label{eq:w}
\end{equation}
where $\mathbf{W_{1}}\in \mathbb{R}^{\frac{C}{r} \times C}$ and $\mathbf{W_{2}}\in \mathbb{R}^{C \times \frac{C}{r}}$ are weights of two fully connected (FC) layers, $r$ is a hyperparameter, $\mathbf{w}$ is the channel-wise weighted vector, $\sigma$ is the sigmoid activation.
The channel-wise feature maps of $\mathbf{X'}$ are activated by $\mathbf{w}$:

\begin{equation}
\mathbf{x''}_{c}=f_{ac}\left(\mathbf{x'}_{c}, w_{c}\right)=w_{c} \cdot \mathbf{x'}_{c},
\end{equation}

where $\mathbf{x''}_{c}$ and $w_{c}$ are the $c$th channel of $\mathbf{X''}\in \mathbb{R}^{T'\times F' \times C}$ and $\mathbf{w}$, respectively. $f_{ac}$ is a channel-wise multiplication function.
Finally, an average pooling layer (Avg Pool) is applied to reduce the size of feature maps. A flowchart of the SE layer is shown in Fig. \ref{fig:SE block}.

A \emph{global average pooling} layer is used after the SE-blocks to get a proper input shape of the Transformer encoder.
We assumed the input of the Transformer encoder as $\mathbf{O} \in \mathbb{R}^{T''\times C}$, where $T''$ is the number of time frames.

\subsection{Transformer encoder}
The second part of the SE-Trans is the Transformer encoder.
The Transformer is a sequence to sequence model which usually contains an encoder and a decoder. 
Considering that our proposed cross-task model is used for classification tasks, we only use the encoder.

\begin{figure}[htbp]
    \centering
	\includegraphics[width=70mm]{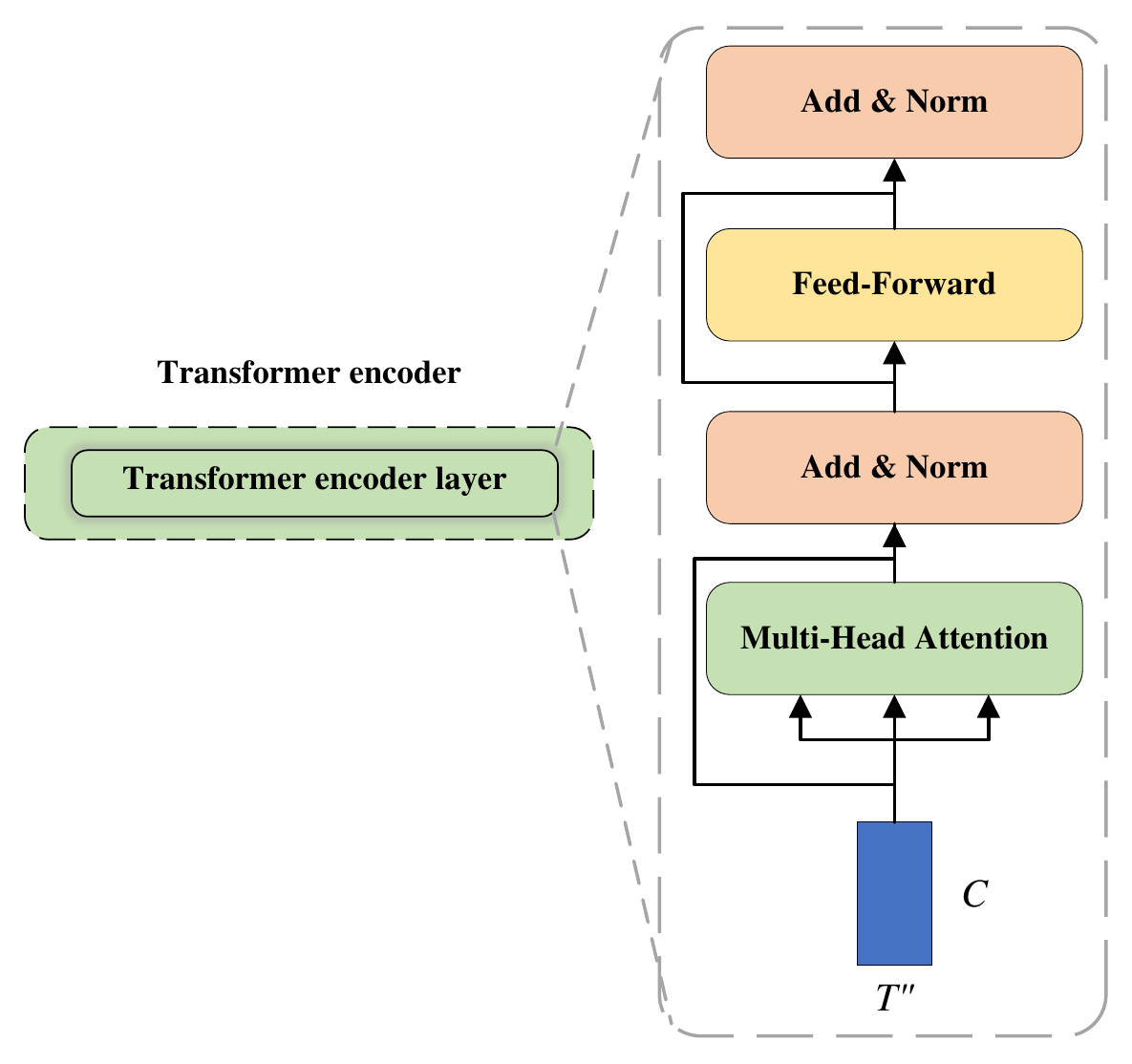}
	\caption{A flowchart of the Transformer encoder layer.}
	\label{fig:Trans}
\end{figure}

In each encoder, there are several encoder layers. 
The $T''$ and $C$ dimension of $\mathbf{O}$ is the sequence length and feature number of the encoder layer, respectively.
Fig. \ref{fig:Trans} shows a flowchart of the encoder layer in the Transformer encoder.
The first part of a Transformer encoder layer is MHSA, which is composed of multiple scaled dot product attention modules.
The input is mapped $h$ times with different, learnable linear projections to get parallel queries $\mathbf{Q}_{i}$, keys $\mathbf{K}_{i}$ and values $\mathbf{V}_{i}$ respectively, and this is formulated as:

\begin{equation}
\begin{aligned}
\mathbf{Q}_{i} &=\mathbf{O}\mathbf{W}^{Q}_{i}\\
\mathbf{K}_{i} &=\mathbf{O}\mathbf{W}^{K}_{i}\\
\mathbf{V}_{i} &=\mathbf{O}\mathbf{W}^{V}_{i},
\end{aligned}
\end{equation}

where $i \in[1, h]$,  
$\mathbf{W}^{Q}_{i},\mathbf{W}^{K}_{i} \in \mathbb{R}^{C \times d_{k}}, \mathbf{W}^{V}_{i} \in \mathbb{R}^{C \times d_{v}}$0

are $i$th linear transformation matrix for mapped queries, keys, and values 

$\mathbf{Q}_{i},\mathbf{K}_{i}\in \mathbb{R}^{T'' \times d_{k}},\mathbf{V}_{i} \in \mathbb{R}^{T'' \times d_{v}}$.

The dot product of the queries $\mathbf{Q}_{i}$ with all keys $\mathbf{K}_{i}$ is then computed, followed by division of a scaling factor $\sqrt{d_{k}}$. 
The softmax operation converts the correlation values $\mathbf{V}_{i}$ to
probabilities which indicates how much the importance of $\mathbf{V}_{i}$ in a time step should be attended.
The output of the scaled dot product attention is computed as a weighted sum of values $\mathbf{V}_{i}$:

\begin{equation}
head_{i}=\operatorname{Attention}\left(\mathbf{Q}_{i}, \mathbf{K}_{i}, \mathbf{V}_{i}\right)=\operatorname{softmax}\left(\frac{\mathbf{Q}_{i} \mathbf{K}^\mathrm{T}_{i}}{\sqrt{d_{k}}}\right) \mathbf{V}_{i}.
\end{equation}

The attentions of all heads are concatenated and linearly projected again to obtain the multi-head output:

\begin{equation}
\mathbf{O}_{MH} =\operatorname{Concat}\left(head_{1}, \cdots, head_{h}\right) \mathbf{W}^{O},
\end{equation}

where $\mathbf{W}^{O}\in \mathbb{R}^{C \times C}$ is a linear transformation matrix.
After that, residual connections and layer normalization (LN) \cite{ba2016layer} are employed:

\begin{equation}
\mathbf{O}_{M}=\operatorname{LN}(\mathbf{O}+\mathbf{O}_{MH}).
\end{equation}

The output is fed into the feed-forward network (FFN) followed by a residual connection and LN to get the final output of the Transformer encoder and that is formulated as:

\begin{equation}
\mathbf{O}^{\prime} = \operatorname{LN}\left(\operatorname{FFN}\left(\mathbf{O}_{M}\right)+\mathbf{O}_{M}\right),
\end{equation}

where $\mathbf{O}^{\prime} \in \mathbb{R}^{T''\times C}$ is the output of the Transformer encoder.

\subsection{Loss functions}
In this section, we describe the loss functions used for ASC, UST, and ASD in the model.

\subsubsection{Loss of acoustic scene classification}
Since the ground truth and prediction of a recording contains only one of the acoustic scene classes, ASC is a multi-class classification task.
We denote the output of SE-Trans as $f(\mathbf{X})=[f(\mathbf{X})_{1},\cdots,f(\mathbf{X})_{N}]$, where $N$ is the number of classes.
The loss function of ASC is categorical cross-entropy loss, defined as:
\begin{equation}
\label{eq:ce}
L_{ASC}(\hat{y}, y)=-\sum_{n=1}^{N} y_{n} \log \hat{y}_{n}
\end{equation}
where $\hat{y}_{n} = \operatorname{softmax}\left(f(\mathbf{X})_{n}\right)=\frac{e^{f(\mathbf{X})_{n}}}{\sum_{n=1}^{N} e^{f(\mathbf{X})_{n}}}$ is the estimated label and $y_{n}$ is the true label of the $n$th class .

\subsubsection{Loss of urban sound tagging}
While for UST, the ground truth and prediction of a sound recording may contain multiple classes. Therefore, it is a multi-label classification task.
The loss function of UST is categorical binary cross-entropy loss, where $f(\mathbf{X})_{n}$ has to pass the $\operatorname{sigmoid}$ activation function to obtain $\hat{y}_{n}$, and the loss is defined as:
\begin{equation}
L_{UST}(\hat{y}, y) =-\sum_{n=1}^{N}\left[y_{n} \log \hat{y}_{n}+\left(1-y_{n}\right) \log \left(1-\hat{y}_{n}\right)\right].
\end{equation}

\subsubsection{Loss of anomalous sound detection}
For ASD in MCM, we adopt a self-supervised learning strategy to train the model to differentiate different machine IDs of one machine type.
Therefore, ASD can be seen as a supervised multi-class classification task.
The loss function is categorical cross-entropy loss and the same as defined in Eq. \ref{eq:ce},
where the $\hat{y}_{n}$ and $y_{n}$ in ASD refer to the estimated and true machine ID, respectively.

\subsection{Data augmentation}
In the proposed cross-task model, we employ FMix as the data augmentation method to improve generalization and prevent overfitting of the neural networks.

We applied FMix to the acoustic features in the training stage, the operations are described as follows.
First, we sample a random complex matrix $\mathbf{Z}\in \mathbb{C}^{T \times F}$, having the same size as $\mathbf{X}$, with both the real and imaginary parts of $\mathbf{Z}$ are independent and Gaussian. 
We then keep the low-frequency components and decay the high-frequency components by applying a low-pass filter on $\mathbf{Z}$.
Next, we perform an inverse Fourier transform on the complex tensor $\mathbf{Z}$ and take the real part to obtain a grey-scale image.
We assign the top $n$ elements of the grey-scale image a value of 1 and the rest a value of 0, therefore, we can obtain a binary mask by the following method:

\begin{equation}
\label{eq6}
mask(\lambda, \mathbf{g})[i,j]=\left\{
\begin{aligned}
1&,\quad\text{if}\quad\mathbf{g}[i,j] \in \text{top}(\lambda TF, \mathbf{g})\\
0&,\quad \text{otherwise},
\end{aligned}
\right.
\end{equation}

where $\mathbf{g}$ refers to the grey-scale image, $mask(\lambda, \mathbf{g})$ refers to the binary mask, and $\lambda \in [0, 1]$ is a hyperparameter. The $\text{top}(n, x)$ function returns a set containing the top $n$ elements of the input $x$, while $\text{top}(\lambda TF, \mathbf{g})$ means that top $\lambda TF$ elements in $\mathbf{g}$ will be selected.
Finally, the mixed feature $\tilde{\mathbf{X}}$ can be obtained from two input features $\mathbf{X}_{i}$ and $\mathbf{X}_{j}$ using the following formulation:
\begin{equation}
\centering
\tilde{\mathbf{X}}=mask(\lambda, \mathbf{g})\odot \mathbf{X}_{i}+(1-mask(\lambda, \mathbf{g}))\odot \mathbf{X}_{j},
\end{equation}
where $\odot$ denotes the Hadamard product.

\section{Experimental procedures}
\label{sec:exp}
We evaluated the proposed cross-task model for all the tasks on the latest dataset of DCASE challenges. 

We first describe details of the cross-task model, and then we
introduce the dataset, experimental setups, baseline models, and evaluation metrics for ASC, UST, and ASD in this section.

\subsection{Cross-task model}

In feature extraction, we used log Mel spectrograms
as the common acoustic features. However, we extract the acoustic features with different configurations for individual tasks in order to make a fair comparison. The details of feature extraction are described in the following subsections.

SE-Trans is used as the main architecture, which contains two SE-blocks and one Transformer encoder.
Each SE block consists of 2 convolutional layers with the same channels and kernel sizes of $3\times3$. 
The number of channels of the first and second SE-block is 64 and 128, respectively. 
An average pooling layer is applied after each SE-block with kernel sizes of $2\times2$. 
Similarly, \emph{AdaptiveAvgPool2d} layer is applied after that to get a proper input size of $16\ (T'')\times 128\ (C)$ for the Transformer encoder. 
We also repeated the experiment with different numbers of heads (4 and 8), layers (1 and 2), and FFN size (16, 32, and 64).
Max aggregation function along time frames along with an FC layer are applied to the output of SE-Trans.
Adam optimizer with a learning rate of 0.001 is used to optimize the loss function.

Besides FMix, two more data augmentation methods were also analyzed.
The first method is SpecAugment, which randomly masks the frequency bins and time frames of the spectrograms.
The second method is mixup, whose operations on training samples are expressed as:
\begin{equation}
\centering
\tilde{x}=\lambda x_{i}+(1-\lambda) x_{j}
\end{equation}
\begin{equation}
\tilde{y}=\lambda y_{i}+(1-\lambda) y_{j},
\end{equation}
where $x_{i}$ and $x_{j}$ are the input features, $y_{i}$ and $y_{j}$ are the corresponding target labels, and $\lambda\in[0,1]$ is a random number drawn from the beta distribution.

\subsection{Acoustic scene classification}
\subsubsection{Dataset}

%\begin{table}[h]
%\caption{Acoustic scenes and cities in the development set of DCASE 2021 Task1 Subtask A.}
%\label{tab:scenes of ASC}
%\centering
%\begin{tabular}{ccc}
%\cline{1-2}
%Acoustic scenes    & Cities    &  \\ \cline{1-2}
%airport            & Amsterdam &  \\
%shopping\_mall     & Barcelona &  \\
%metro station      & Helsinki  &  \\
%street\_pedestrain & Lisbo     &  \\
%public\_square     & London    &  \\
%%street\_traffic    & Lyon      &  \\
%tram               & Madrid    &  \\
%bus                & Milan     &  \\
%metro              & Prague    &  \\
%park               & Paris     &  \\
%                   & Stockholm &  \\
%                   & Vienna    &  \\ \cline{1-2}
%\end{tabular}
%\end{table}

The dataset used for ASC is the development set of DCASE 2021 Task1 Subtask A \cite{Heittola2020}. 
The organizers used different devices to simultaneously capture audio in 10 acoustic scenes, which are airport, shopping mall, metro station, street pedestrian, public square, street traffic, tram, bus, metro, and park.
The development set contains data from 9 devices: A, B, C (3 real devices), and S1-S6 (6 simulated devices).
The total number of recordings in the development set is 16,930.
The dataset is divided into a training set, which contains 13,962 recordings, and a testing set, which contains 2,968 recordings. 
Complete details of the development set are shown in Table \ref{tab:Statistics of ASC}.

\begin{table}[h]
\caption{Statistics of the development set of ASC.}
\label{tab:Statistics of ASC}
\centering
\renewcommand\arraystretch{1.25}	
\begin{tabular}{cccc}
\hline\hline
\multicolumn{2}{c}{Devices} & \multicolumn{2}{c}{Development set} \\\cmidrule(r){1-2} \cmidrule(r){3-4}
Name        & Type          & Training set      & Testing set      \\ \hline
A\&B\&C     & real          & 10,215\&749\&748   & 330\&329\&329   \\
S1-S6       & simulated     & 2,250              & 1,980            \\ \hline
Total       &               & 13,962             & 2,968            \\ \hline\hline
\end{tabular}
\end{table}

\subsubsection{Experimental setups}
\label{sec: ASC Experimental setups}
We used log Mel spectrograms as the input features. 
First, all recordings were resampled to 44,100 Hz. 
The short-time Fourier transform (STFT) with a Hanning window of 40 ms and a hop size of 20 ms was used
to extract the spectrogram.
We applied 40 Mel-filter bands on the spectrograms followed by a logarithmic operation to calculate log Mel spectrograms. 
Each log Mel spectrogram has a shape of $500\times40$, where 500 is the number of time frames and 40 is the number of frequency bins.

\subsubsection{Baseline models}
\label{sec: ASC Baseline systems}

We used three models as baseline in the experiments.
The first are official baseline models, which are provided by the task organizers.
The official baseline model for ASC is a 3-layer CNN model, which consists of 16, 16, and 32 feature maps for each convolutional layer, respectively \cite{Mart2021lowcomplexity}.

The second type is CNNs-based models, which are proposed by Kong et al. in the study of cross-task models \cite{kong2019cross}.
The CNN5 consists of 4 convolutional layers with a kernel size of $5\times5$ and feature maps of 128, 128, 256, and 512. 
The CNN9 consists of 4 convolutional blocks, where feature maps of 64, 128, 256, and 512 and a kernel size of $3\times3$ are applied.
The CNN13 consists of 6 convolutional blocks, where feature maps of 64, 128, 256, 512, 1024, and 2048 and a kernel size of $3\times3$ are applied.
For all architectures, BN, ReLU and average pooling layers with a size of $2\times2$ are applied. 

The third is a CRNN-based model, which aims for audio tagging task and achieves great performance \cite{kong2020sound}.
This model uses the same architecture as CNN9, where the frequency axis of the output from the last convolutional layer is averaged. 
Then a bidirectional gated recurrent unit (biGRU) and time distributed fully connected layer is applied to predict the presence of sound classes.
Mixup is exploited during the training stage and the model is named CNN-biGRU-Avg.
Finally, our proposed cross-task model is compared with 5 baseline models: official baseline models, CNN5, CNN9, CNN13, and CNN-biGRU-Avg.

\subsubsection{Evaluation metrics}
In order to evaluate the performance of the models, we first compute the accuracy (ACC), precision (P) and recall (R) of class $n$ as follows:
\begin{equation}
\label{eq:acc}
\text{ACC}_{n}=\frac{\text{TP}}{\text{TP}+\text{FP}}
\end{equation}
\begin{equation}
\label{eq:precision}
\text{P}_{n}=\frac{\text{TP}}{\text{TP}+\text{FP}}
\end{equation}
\begin{equation}
\label{eq:recall}
\text{R}_{n}=\frac{\text{TP}}{\text{TP}+\text{FN}},
\end{equation}
where TP, FP, and FN are the number of true positive, false positive, and false negative samples, respectively.
Then, the macro-average accuracy (macro-ACC) is defined as:
\begin{equation}
\label{eq:macro-ACC}
\text{macro-ACC}=\frac{1}{N}\sum_{n=1}^{N}\text{ACC}_{n}.
\end{equation}

\subsection{Urban sound tagging}
\subsubsection{Dataset}
The dataset used for UST is the development set of DCASE 2020 Task5, named sounds of New York City urban sound tagging (SONYC-UST-V2) \cite{cartwright2020sonyc}.
SONYC-UST-V2 consists of 8 coarse-level and 23 fine-level urban sound categories, where the provided audio has been acquired using the SONYC acoustic sensor network for urban noise pollution monitoring. 
50 sensors have been laid out in different areas of New York City, and these sensors have collected a large number of audio data. 

The development set is grouped into disjoint sets, containing a training set (13,538 recordings) and a testing set (4,308 recordings). 
All recordings are of 10 seconds and the annotation for each recording contains reference labels, which are annotated by at least one annotator, and spatiotemporal context. 
The fine-level reference labels and spatiotemporal context are not used in our experiments, only 8 coarse-level classes are used for UST, including engine, machinery impact (M/C), non-machinery impact (non-M/C), powered saw (saw), alert signal (alert), music, human voice (human), and dog.
In addition, the official baseline used the verified annotations of the validate split, which leads to a validate split of 538 recordings.

% \begin{table}[h]
% 	\centering
% 	\setlength{\tabcolsep}{9mm}
% 	\caption{Coarse-level classes abbreviations of SONYC-UST-V2.}
% 	\label{tab: UST abbr}
	
% 	\begin{tabular}{cc}
% 		\hline
% 		Classes       & Abbreviations \\ \hline
% 		Engine               & -       \\
% 		Machinery impact     & M/C           \\
% 		Non-machinery impact & Non-M/C       \\
% 		Powered saw          & Saw           \\
% 		Alert signal         & Alert         \\
% 		Music                & -         \\
% 		Human voice          & Human         \\
% 		Dog                  & -           \\ \hline
% 	\end{tabular}
% \end{table}

\subsubsection{Experimental setups}
We resampled all recordings to 20,480 Hz and applied STFT on them with a Hanning window of 1,024 samples and a hop size of 512 samples.
The log Mel spectrograms with 64 Mel-filter bands are used as input features, each having a shape of $401\times64$.
The other setups are the same as described in Sec. \ref{sec: ASC Experimental setups}.

\subsubsection{Baseline models}
Three types of models were used as baseline models for UST as well.
The official baseline model of DCASE 2020 Task5 uses a multi-layer perceptron model, which consists of one hidden layer of size 128 and an autopool layer \cite{cartwright2019sonyc}. 
OpenL3 embeddings are taken as the input using a window size and hop size of 1 second.
The model was trained using stochastic gradient descent to minimize binary cross-entropy loss under L2 regularization.  
The remaining two types of models contain CNN5, CNN9, CNN13, and CNN-biGRU-Avg, which have the same configurations as described in Sec. \ref{sec: ASC Baseline systems}.

\subsubsection{Evaluation metrics}
We use the macro area under the precision-recall curve (macro-AUPRC) as the classification metric. 
The macro-precision (macro-P) and macro-recall (macro-R) are defined as:
\begin{equation}
\label{eq:macroP}
\text{macro-P}=\frac{1}{N}\sum_{n=1}^{N}\text{P}_{n}
\end{equation}
\begin{equation}
\label{eq:macroP}
\text{macro-R}=\frac{1}{N}\sum_{n=1}^{N}\text{R}_{n},
\end{equation}
we changed the threshold from 0 to 1 to compute different macro-P and macro-R, and calculate the area under the P-R curve to get macro-AUPRC.
Moreover, micro-AUPRC and micro-F1 score are used as additional metrics.

\subsection{Anomalous sound detection}
\subsubsection{Dataset}
The dataset used for ASD is the development set of DCASE 2021 Task2 \cite{Tanabe_arXiv2021_01, harada2021toyadmos2}, consisting of normal and anomalous sounds of 7 types of machines. 
Each recording is single-channel 10-second audio, which is recorded in a real environment. 
Two important pieces of information about the machines are provided, i.e., machine type and machine ID.
Machine ID is the identifier of the same type of machine, which is provided in the dataset by different section information, such as section 00, 01, and 02.
Machine type means the kind of machine, which can be one of the following: toyCar, toyTrain, fan, gearbox, pump, slide rail (slider), and valve.

\begin{figure}[htbp]
	\centering
	\includegraphics[width=7cm]{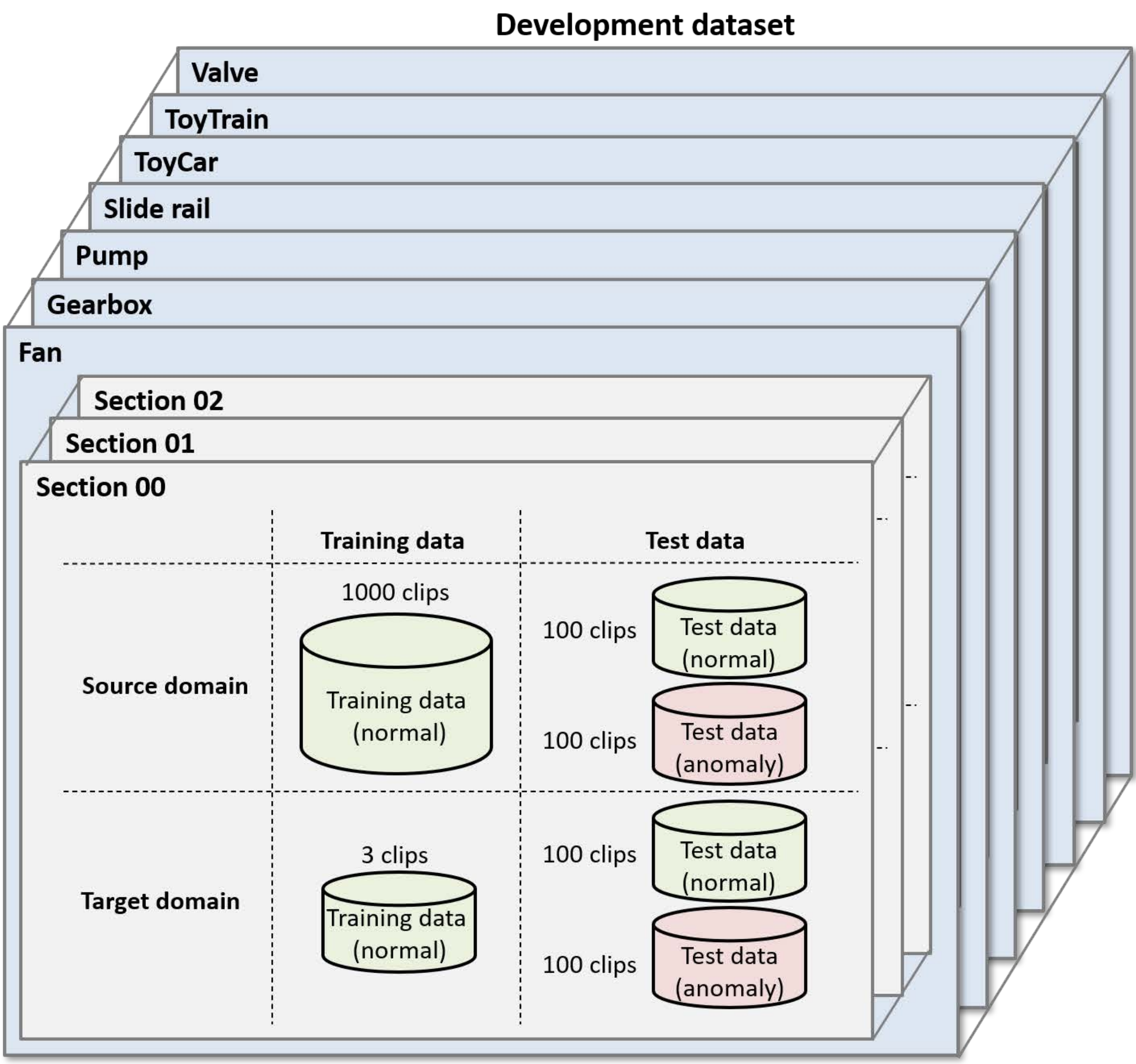}
	\caption{An overview of the development set of ASD \cite{dcase2021task2url}.}
	\label{fig:task2 datasets}
\end{figure}

This dataset consists of three sections for each machine type, and each section comprises training and test data.
In the source domain of each section, there are around 1,000 clips of normal sound recordings for training, and around 100 clips each of normal and anomalous sound recordings for test.
While in the target domain, only 3 clips of normal sound recordings are provided for training, and around 100 clips each of normal and anomalous sounds are provided for test.
Fig. \ref{fig:task2 datasets} shows an overview of the development set of ASD.

\subsubsection{Experimental setups}
Log Mel spectrograms are used as input features for ASD models.
The recordings are loaded with the default sample rate of 16,000Hz.
We applied STFT with a Hanning window size of 1,024 and a hop length of 512 samples.
Mel-filters with bands of 128 are used to generate log Mel spectrograms.
In addition, the input features are obtained by concatenating consecutive 64 frames of the log Mel spectrograms. 
The final shape of the input to the network is $64\times128$.
The other setups are the same as described in Sec. \ref{sec: ASC Experimental setups}.

\subsubsection{Baseline models}
There are also three types of baseline models for ASD.
The official baseline of DCASE 2021 Task2 is a MobileNetV2-based model \cite{Kawaguchi_arXiv2021_01}.
This model is trained in a self-supervised manner using the IDs of the machines.
Besides the official baseline model, the remaining two types of models (CNN5, CNN9, CNN13, and CNN-biGRU-Avg) are the same as described in Sec. \ref{sec: ASC Baseline systems}

\subsubsection{Evaluation metrics}
This task is evaluated with the area under the curve (AUC) of the receiver operating characteristic (ROC).
We first define the anomaly score $A(\mathbf{X})$ as:

\begin{equation}
\begin{aligned}
A(\mathbf{X})&=\operatorname{log}(\frac{1-f(\mathbf{X})}{f(\mathbf{X})})\\
&=\frac{1}{B}\sum_{b=1}^{B}\operatorname{log}(\frac{1-f(\mathbf{X}_{b})}{f(\mathbf{X}_{b})})\\
&=\frac{1}{B}\sum_{b=1}^{B}\operatorname{log}(\frac{1-\hat{y}_{b}}{\hat{y}_{b}}),
\end{aligned}
\end{equation}

where $B$ is the number of the input features extracted from log Mel spectrograms by shifting a context window by $8$ frames,  and $\hat{y}_{b}$ is the softmax output of the network.
The AUC can be defined as:

\begin{equation}
\text{AUC} = \frac{1}{N_{-}N_{+}} \sum_{i=1}^{N_{-}} \sum_{j=1}^{N_{+}} \mathcal{H} (A(\mathbf{X}_{j}^{+}) - A(\mathbf{X}_{i}^{-})),
\end{equation}

where $\mathbf{X}_{j}^{+}$ and $\mathbf{X}_{i}^{-}$ are normal and anomalous test input features, $N_{+}$ and $N_{-}$ are the number of normal and anomalous test samples, respectively.
And $\mathcal{H}(x)$ returns 1 when $x>0$ and 0 otherwise. 
Moreover, the partial-AUC (pAUC) is used as an additional metric, which is calculated from a portion of ROC over the pre-specified range of interest. In this task, the pAUC is calculated as the AUC over a low false-positive-rate (FPR) range [0,p], where we will use $p=0.1$.

\section{Results and discussions}
\label{sec:results}
In this section, we demonstrate the results of the experiments and give further discussions about the models from two aspects: cross-task, where the generality is analyzed, and subtask, where the individuality is analyzed.
In addition, we visualize the attention masks of the proposed SE-Trans to study the effectiveness of the attention mechanism.

\subsection{Cross-task}
For the cross-task aspect, we compare the general performance of the proposed cross-task model with other baseline models, and further investigate the importance of SE, Transformer modules, and data augmentation methods in our model.
\subsubsection{Comparison of different models}

\begin{table*}[t]
\caption{Cross-task performance comparison with different models.}
\label{tab:Results of cross-task methods}
\centering
\renewcommand\arraystretch{1.25}
\begin{tabular}{ccccccc}
\hline\hline
  \multirow{2}{*}{Models} & ASC            & \multicolumn{3}{c}{UST}                          & \multicolumn{2}{c}{ASD}         \\ \cmidrule(r){2-2} \cmidrule(r){3-5} \cmidrule(r){6-7}
                        & macro-ACC      & macro-AUPRC    & micro-AUPRC    & micro-F1       & AUC            & pAUC           \\ \hline
Official baseline \cite{Mart2021lowcomplexity} \cite{cartwright2019sonyc} \cite{Kawaguchi_arXiv2021_01}               & 0.477          & 0.632          & 0.835          & 0.739          & 0.622          & 0.574          \\
CNN5 \cite{kong2019cross}                    & 0.525          & 0.683          & 0.860          & 0.766          & 0.686          & 0.613          \\
CNN9 \cite{kong2019cross}                     & 0.550          & 0.678          & 0.862          & 0.767          & 0.703          & 0.616          \\
CNN13 \cite{kong2019cross}                    & 0.475          & 0.675          & 0.851          & 0.736          & 0.708          & 0.612          \\
CNN-biGRU-Avg \cite{kong2020sound}        & 0.572          & 0.684          & 0.845          & 0.763          & 0.711          & 0.617          \\
Proposed                & \textbf{0.633} & \textbf{0.727} & \textbf{0.872} & \textbf{0.785} & \textbf{0.751} & \textbf{0.632} \\ \hline\hline
\end{tabular}
\end{table*}

\begin{figure}[t]
	\centering
	\includegraphics[scale=0.55]{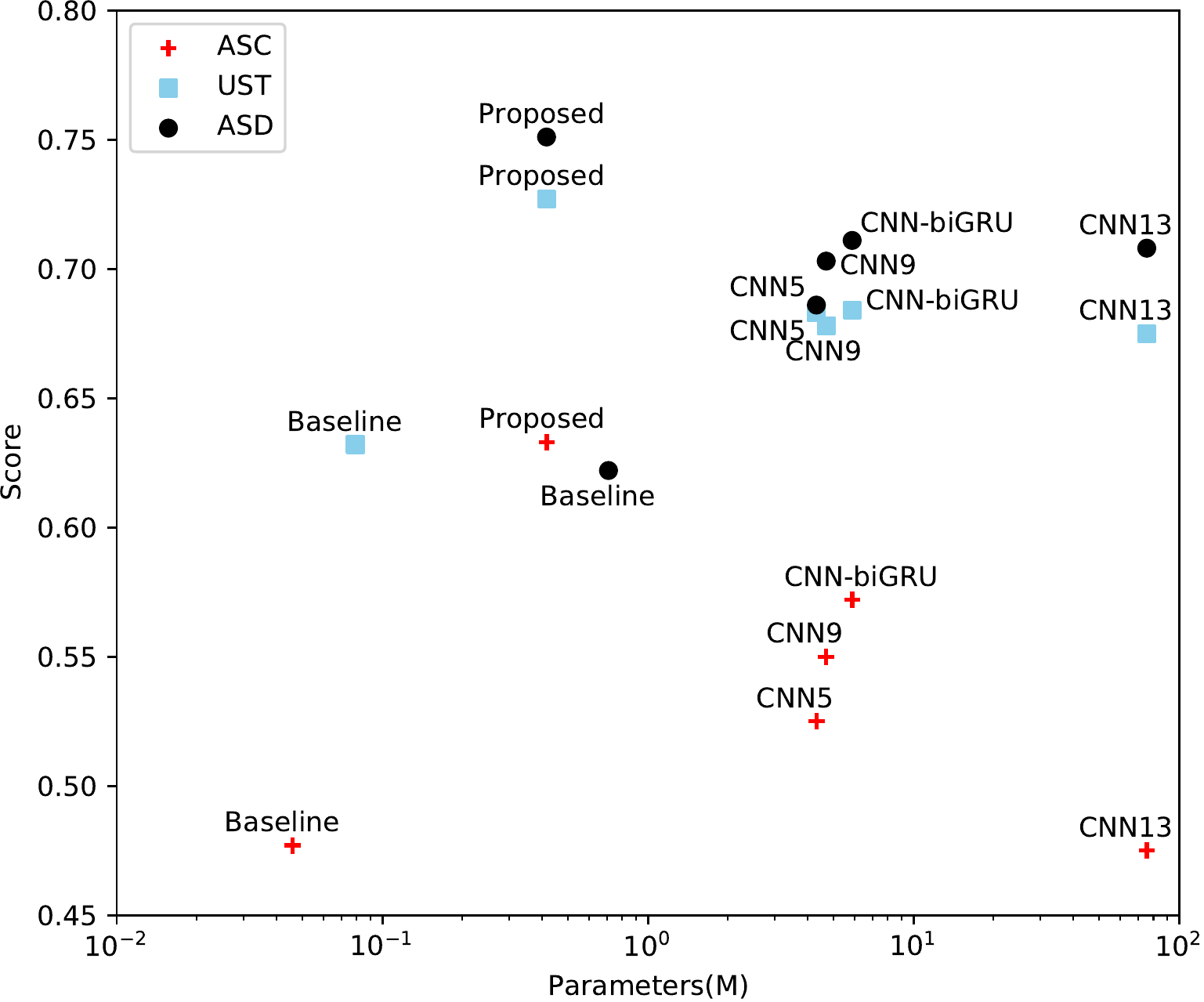}
	\caption{Parameters versus scores of different models.
	Official baseline (Baseline), CNN-biGRU-Avg (CNN-biGRU).}
	\label{fig:Params_and_scores}
\end{figure}

In this section, we compare the proposed cross-task model with state-of-the-art methods on different tasks, i.e., ASC, UST, and ASD. 
Table \ref{tab:Results of cross-task methods} shows that the proposed cross-task model surpasses the performance of official baselines, CNN5, CNN9, CNN13, and CNN-biGRU-Avg on all evaluation metrics.
For ASC, the proposed model achieves a macro-ACC of 0.633; for UST, our model achieves a macro-AUPRC of 0.727, a micro-AUPRC of 0.872, and a micro-F1 of 0.785; for ASD, the proposed model achieves AUC and pAUC of 0.751 and 0.632, respectively.

In real applications, the number of parameters and performance has to be considered and there must be a trade-off between them. 
Therefore, we further analyzed the model complexity of the aforementioned models.
Fig. \ref{fig:Params_and_scores} shows the parameters versus scores of different models for ASC, UST, and ASD, while the score used for comparison on each task is macro-ACC, macro-AUPRC, and AUC.
As shown in the figure, CNN-biGRU-Avg performs better than the official baseline models, CNN5, CNN9, and CNN13 for all tasks, containing about 5.8 million parameters.
The proposed cross-task model has achieved significant improvement and outperforms the models under comparison for all tasks.
However, our model contains around 0.4 million parameters, which are only 7\% of the model complexity of CNN-biGRU-Avg.

\subsubsection{Analysis of SE and Transformer modules}

\label{Analysis of Trans and SE}
To further study the effectiveness of the Transformer encoder and SE modules, ablation experiments have been performed.
Table \ref{tab:Ablation of Trans and SE layer} compares three different architectures with different numbers of CNN channels and blocks, and macro-ACC, macro-AUPRC, and AUC are implemented as evaluation measurements for ASC, UST, and ASD.
CNN9 contains 4 CNN blocks with different numbers of channels, while CNN4-Trans contains 2 CNN blocks followed by a Transformer encoder.
Then each CNN block in CNN4-Trans is replaced by an SE-block to get SE-Trans.
For ASC, the CNN4-Trans achieves a macro-ACC of 0.589, outperforming the CNN9 by 7\%.
The SE-Trans further improves the macro-ACC of CNN4-Trans from 0.589 to 0.609.
For UST, the  performance of CNN4-Trans remains the same as macro-AUPRC of 0.678 with CNN9.
SE-Trans achieves a macro-AUPRC of 0.699, outperforming the other networks.
For ASD, the CNN4-Trans obtains an improvement of 0.008 in AUC and the SE-Trans obtains a further improvement of 0.009 in AUC.
To summarize, both of the SE and Transformer modules improve the performance.
The main reasons are that the channel-wise attention in SE layers and the MHSA in Transformer encoder can improve the performance of classifying similar acoustic scenes, tagging various urban sounds and detecting anomalous machines.
Nevertheless, the contribution of these two modules varies for different tasks. 
The CNN4-Trans benefits much from the Transformer encoder for ASC, the SE-Trans achieves greater performance compared to CNN4-Trans for UST, both the Transformer encoder and SE layers can greatly improve the ASD performance.

\begin{table}[t]
\caption{Results of ablation experiments of SE and Transformer modules.}
\label{tab:Ablation of Trans and SE layer}
\centering
\setlength\tabcolsep{4pt}
\renewcommand\arraystretch{1.25}	
\begin{tabular}{ccccc}
\hline\hline
\multirow{2}{*}{Network} & \multirow{2}{*}{Channels}   & ASC       & UST         & ASD   \\ \cmidrule(r){3-3} \cmidrule(r){4-4} \cmidrule(r){5-5}
                         &                                      & macro-ACC & macro-AUPRC & AUC   \\  \hline
CNN9                     & 64/128/256/512                  & 0.550     & 0.678       & 0.703 \\
CNN4-Trans               & 64/128                      & 0.589     & 0.678       & 0.712 \\
SE-Trans                 & 64/128                          & \textbf{0.609}     & \textbf{0.699}       & \textbf{0.721} \\ \hline\hline
\end{tabular}
\end{table}

\subsubsection{Setups of Transformer encoder}
We explored the importance of different numbers of layers, heads, and nodes of FFN in the Transformer encoder. 
Table \ref{tab:params of trans} shows the performance of SE-Trans for all the tasks using different numbers of head, layers, and nodes of FFN.
For ASC, the best performance is achieved by the configuration of 1 layer, 8 heads, and 16 (or 32) nodes of FFN.
For UST, the best number of layers, heads, and nodes of FFN is 1, 8, and 32, respectively.
For ASD, the same configuration of 1 layer, 8 heads, and 32 nodes of FFN achieves the best AUC in comparison.
Considering the performance of different configurations across these tasks, more layers do not achieve any benefit, and the number of heads and nodes of FFN should be chosen carefully.
The combination of 1 layer, 8 heads, and 32 nodes of FFN can be considered as a proper choice for the proposed cross-task model.

\begin{table}[t]
\caption{Performance of SE-Trans using different numbers of head, layers and nodes of FFN.}
\label{tab:params of trans}
\centering
\renewcommand\arraystretch{1.25}
\begin{tabular}{cccccc}
\hline\hline
\multirow{2}{*}{Layers} & \multirow{2}{*}{Heads} & \multirow{2}{*}{FFN} & ASC            & UST            & ASD            \\ \cmidrule(r){4-4} \cmidrule(r){5-5} \cmidrule(r){6-6}
                        &                        &                      & macro-ACC      & macro-AUPRC    & AUC            \\ \hline
1                       & 8                      & 32                   & \textbf{0.609} & \textbf{0.699} & \textbf{0.721} \\
2                       & 8                      & 32                   & 0.592          & 0.683          & 0.713          \\
1                       & 4                      & 32                   & 0.602          & 0.697          & 0.720          \\
1                       & 8                      & 16                   & \textbf{0.609} & 0.682          & 0.718          \\
1                       & 8                      & 64                   & 0.607          & 0.687          & 0.710          \\ \hline\hline
\end{tabular}
\end{table}

\begin{figure*}[t]
\centering
\includegraphics[scale=0.45]{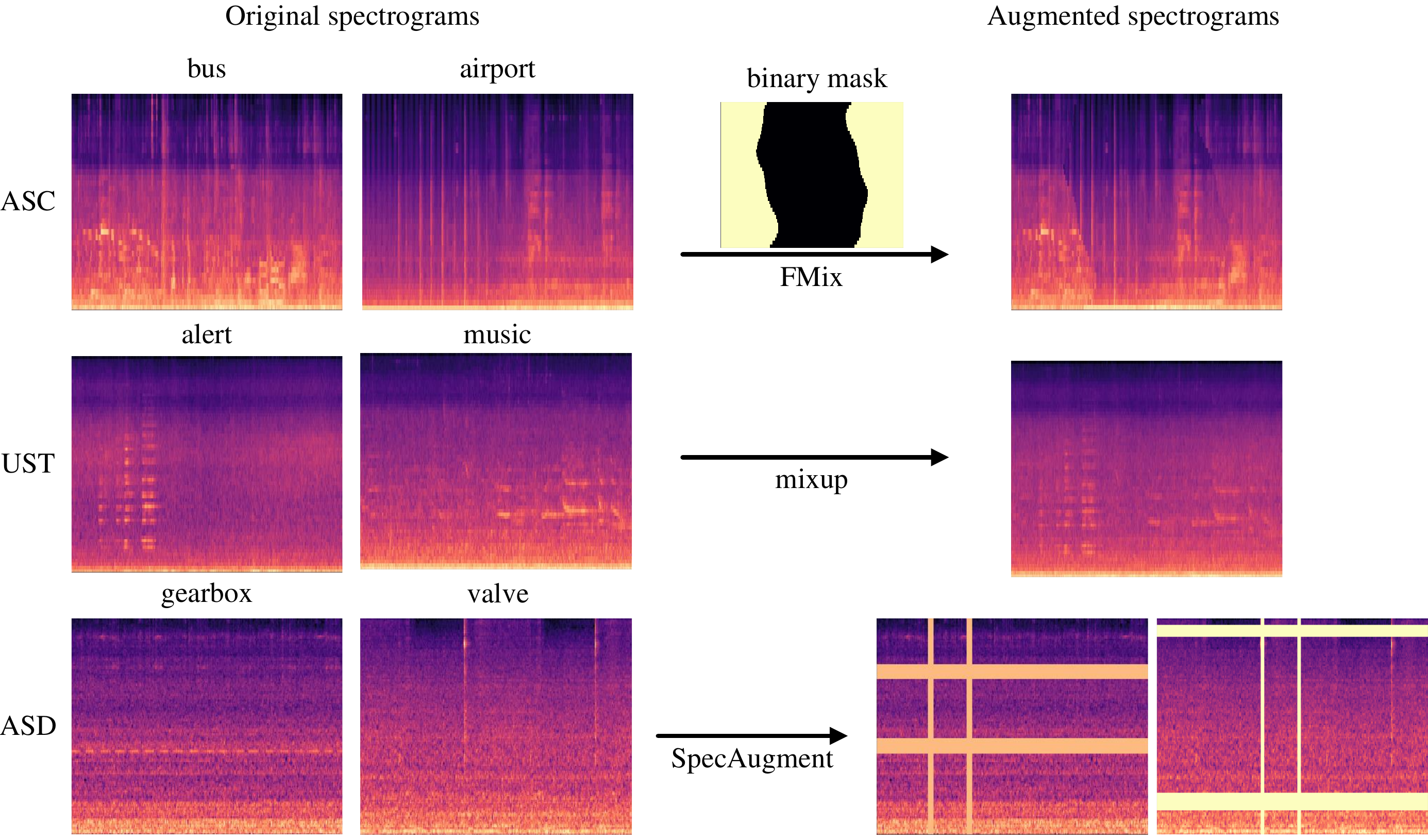}
\caption{Examples of spectrograms augmented by SpecAugment, mixup and FMix for ASC, UST and ASD.}
\label{fig:data_aug_pics}
\end{figure*}

\subsubsection{Results of data augmentation methods}\label{sec:Data augmentation}

In this section, we investigate the performance of using different data augmentation methods: SpecAugment, mixup, and FMix.
Comparative experiments are conducted on the proposed SE-Trans, and the experimental results are shown in Table \ref{tab:Ablation of data augmentation}.
For ASC, FMix improves the macro-ACC from 0.609 to 0.633, outperforming SpecAugment and mixup.
For UST, FMix and mixup achieves an improvement of 0.022 and 0.028 on macro-AUPRC respectively, but applying SpecAugment does not improve the performance of UST.
For ASD, FMix has significantly improved the AUC from 0.721 to 0.751, outperforming the comparative methods by a large margin.
However, SpecAugment brings a slight reduction on the AUC score.
We assume that SpecAugment is not suitable for different ESR tasks, because some key parts of the time-frequency features are randomly masked.

To further analyze, the examples of spectrograms augmented by SpecAugment, mixup and FMix for the tasks are shown in Fig. \ref{fig:data_aug_pics}. 
In the left panel in Fig. \ref{fig:data_aug_pics}, the spectrograms are examples of ASC, UST, and ASD, and the augmented spectrograms are listed in the right panel.
FMix mixes two spectrograms by applying an irregular binary mask, which can generate a more complex spectrogram than mixup.
Therefore, the model can learn robust acoustic features from the irregular areas on the complex spectrograms.
SpecAugment partially masks time-frequency information by randomly setting columns and rows of the spectrograms to zero.
This is effective for speech recognition, because the voice features can not be fully covered.
However, some key acoustic information can be covered for ESR tasks.
Considering the characteristics of environmental sound, SpecAugment should be carefully applied to ESR tasks.

\begin{table}[t]
\caption{Results of ablation experiments of data augmentation methods.}
\label{tab:Ablation of data augmentation}
\centering
\renewcommand\arraystretch{1.25}
\begin{tabular}{cccc}
\hline\hline
\multirow{2}{*}{Data aug.} & ASC            & UST            & ASD            \\ \cmidrule(r){2-2} \cmidrule(r){3-3} \cmidrule(r){4-4}
                          & macro-ACC      & macro-AUPRC    & AUC            \\ \hline
-                         & 0.609          & 0.699          & 0.721          \\
SpecAugment               & 0.607          & 0.701          & 0.717          \\
mixup                     & 0.613          & 0.721          & 0.727          \\
FMix                      & \textbf{0.633} & \textbf{0.727} & \textbf{0.751} \\ \hline\hline
\end{tabular}
\end{table}

\subsection{Subtask}
For the subtask aspect, we further illustrate the performance of the proposed model on individual tasks.
\subsubsection{Analysis of acoustic scene classification}
\begin{figure*}[htbp]
\centering
\includegraphics[scale=0.44]{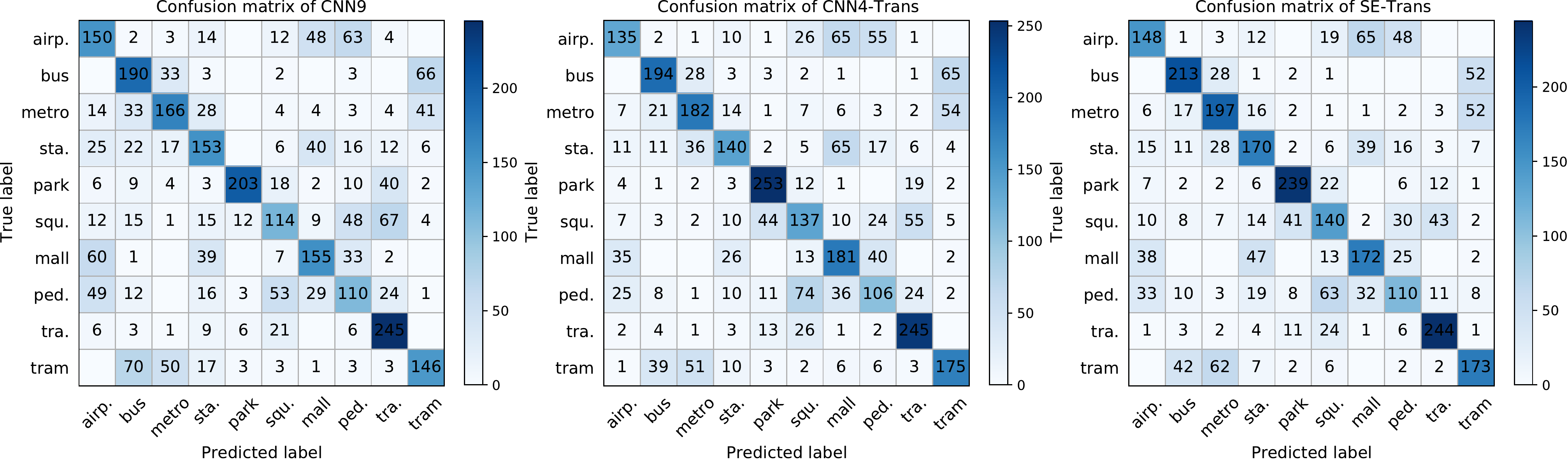}
\caption{Confusion matrices of CNN9, CNN4-Trans and SE-Trans for ASC. Airport (airp.), shopping mall (mall), metro station (sta.), street pedestrian (ped.), public square (squ.), street traffic (tra.).}\label{fig:Confusion matrix}
\end{figure*}

To further study the aforementioned contribution of Transformer and SE modules for ASC in Sec. \ref{Analysis of Trans and SE}, the confusion matrices achieved by CNN9, CNN4-Trans, and SE-Trans are shown in Fig. \ref{fig:Confusion matrix}.
CNN4-Trans and SE-Trans perform better than CNN9 for most of the classes, and the contribution of Transformer and SE modules can be illustrated by analyzing similar acoustic scenes. 
For example, the similar transportation scenes: bus, metro, and tram, are better classified; the performance of recognizing two open and outdoor scenes: park and public square, has been improved as well.
The above observations verify that the attention mechanism in the Transformer encoder and SE layers can improve the performance of classifying similar scenes. 
Furthermore, some classes are not truly predicted, such as airport and street pedestrian.
Since these scenes contain strong background noise and interference of human speech.

%\begin{figure}[htbp]
%\begin{minipage}[t]{0.6\linewidth}
%\centering
%\includegraphics[scale=0.22]{pics/CM_CNN9.png}
%\label{subfig:task1A CNN9}
%\end{minipage}
%\begin{minipage}[t]{0.6\linewidth}
%\centering
%\includegraphics[scale=0.22]{pics/CM_CNN4-Trans.png}
%\label{subfig:task1A CNN4Trans}
%\end{minipage}\\
%\begin{minipage}[t]{1.2\linewidth}
%\centering
%\includegraphics[scale=0.22]{pics/CM_SE-Trans_1.png}
%\label{subfig:task1A SETrans}
%\end{minipage}
%\caption{Confusion matrices of ASC achieved by CNN9, CNN4-Trans and SE-Trans.}\label{fig:Confusion matrix}
%\end{figure} 

\subsubsection{Analysis of urban sound tagging}
\begin{figure}[t]
	\centering
	\includegraphics[scale=0.4]{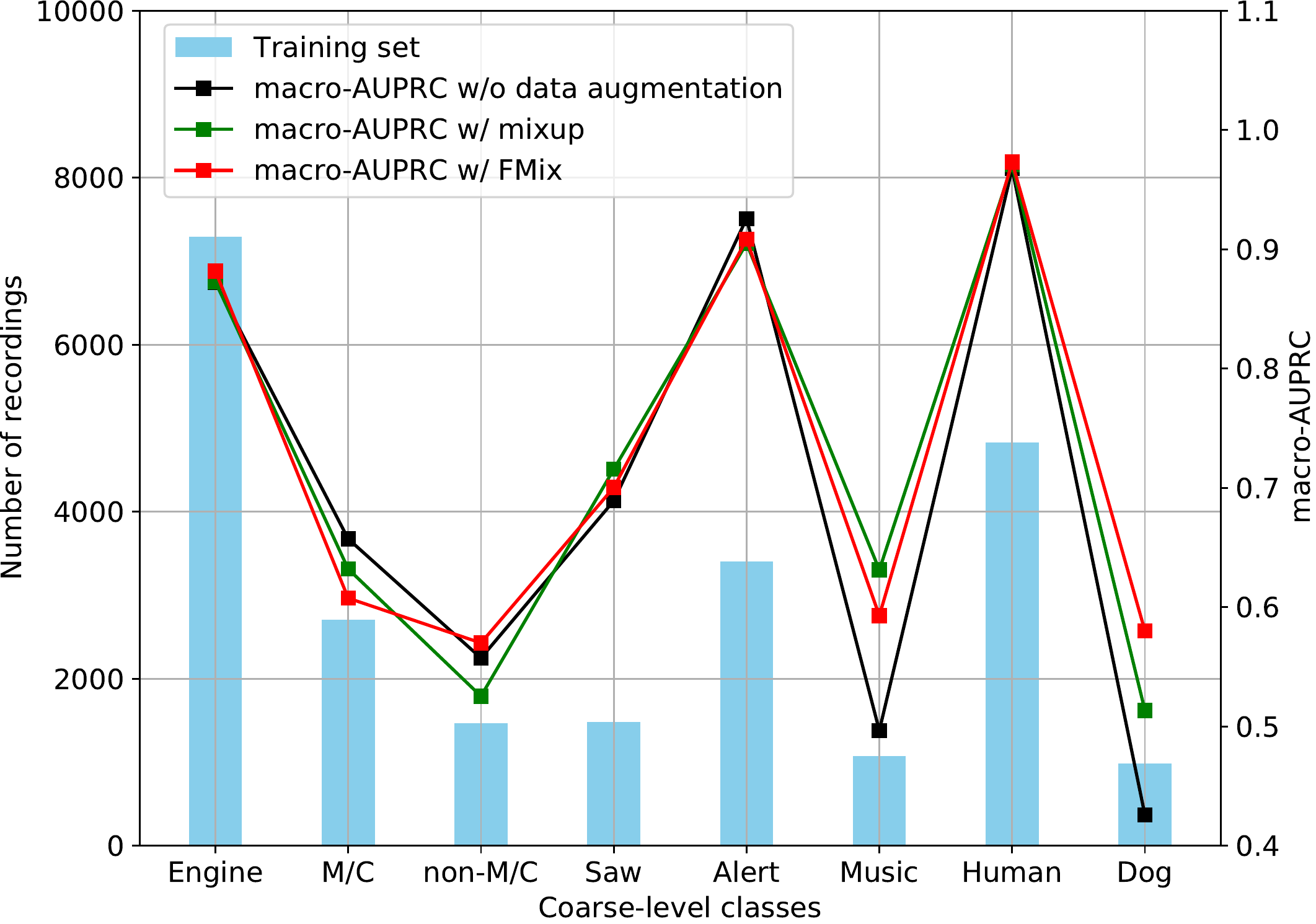}
	\caption{The number of recordings of 8 coarse-level urban sound categories and class-wise macro-AUPRC with different data augmentation methods. With (w/), without (w/o).}
	\label{fig:UST with diff aug}
\end{figure}

We investigate the class-wise performance of UST using different MSDA methods in this section.
First of all, Fig. \ref{fig:UST with diff aug} shows the number of recordings of 8 coarse-level urban sound categories.
The SONYC-UST-V2 dataset is imbalanced and the amounts of non-M/C, saw, music, and dog are less than other categories.
These urban sound classes benefit from applying mixup or FMix, especially for music and dog.
We assume that MSDA methods indeed generate more effective samples, which can help the model learn robust acoustic features and improve the classification performance especially for classes with fewer samples.

\subsubsection{Analysis of anomalous sound detection}
\begin{figure*}[htbp]
\centering
\includegraphics[scale=0.45]{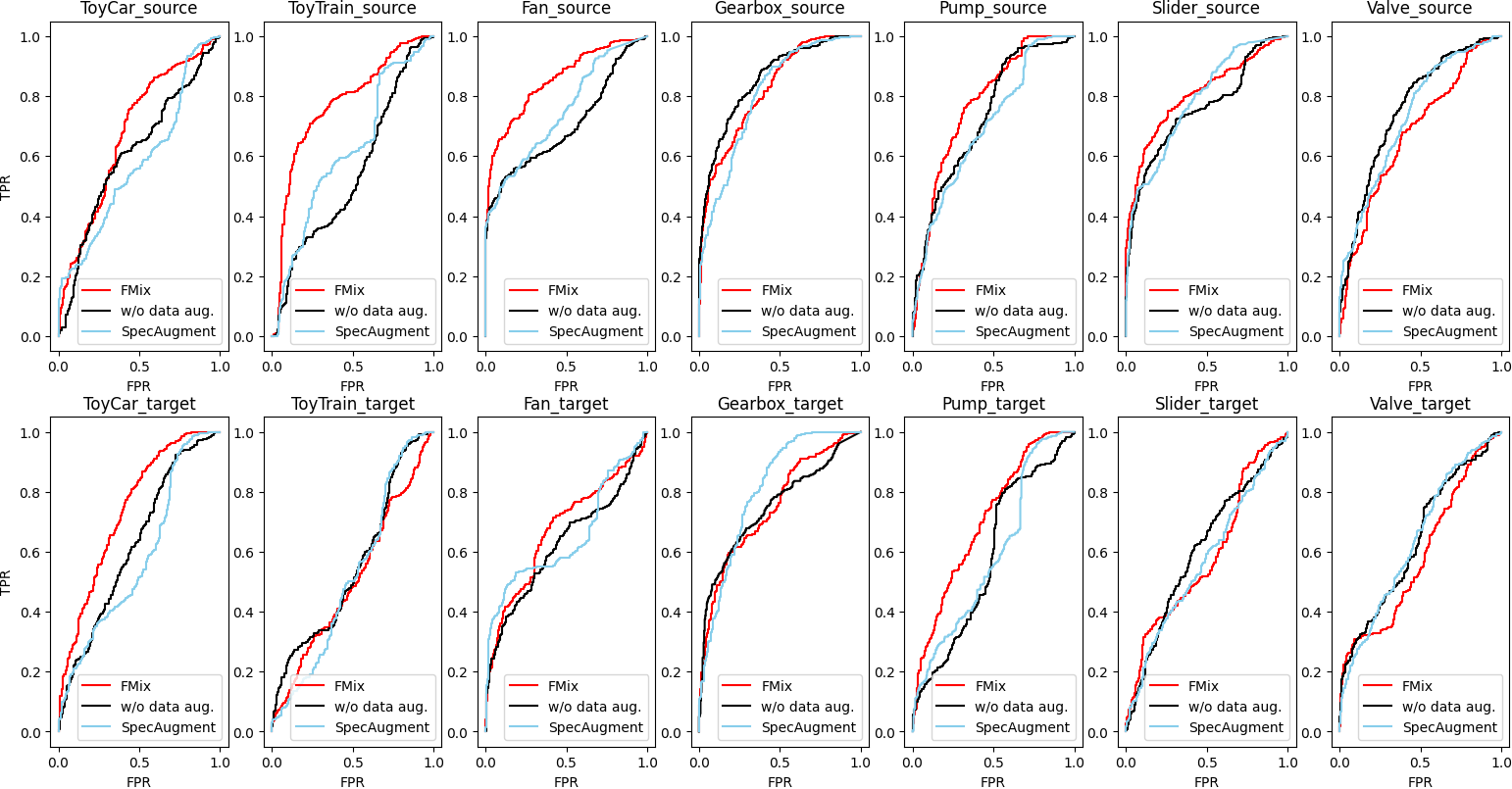}
\caption{ROC curves of different machine types with different data augmentation methods. Without data augmentation (w/o data aug.), source domain (source), target domain (target).}
\label{fig:ASD AUC curves}
\end{figure*}

In Sec. \ref{sec:Data augmentation}, we have shown some examples of spectrograms and analyzed the cross-task performance of different data augmentation methods.
In this section, we compare the performance of ASD using FMix and SpecAugment, and the ROC curves of different machine types are illustrated in Fig. \ref{fig:ASD AUC curves}.
As shown in Fig. \ref{fig:ASD AUC curves}, FMix performs the best for most of the machine types, where the anomalous sounds of toyCar, toyTrain, and fan can be correctly detected.
These sounds have continuous acoustic characteristics, indicating that FMix can improve the performance of these machines.
The performance achieved by SpecAugment is relatively poor on many machine types.
This can verify the assumption in Sec. \ref{sec:Data augmentation} that some key acoustic features on the spectrograms are randomly covered while applying SpecAugment.

\subsection{Visualization}

\begin{figure}[htbp]
\centering
\includegraphics[scale=0.45]{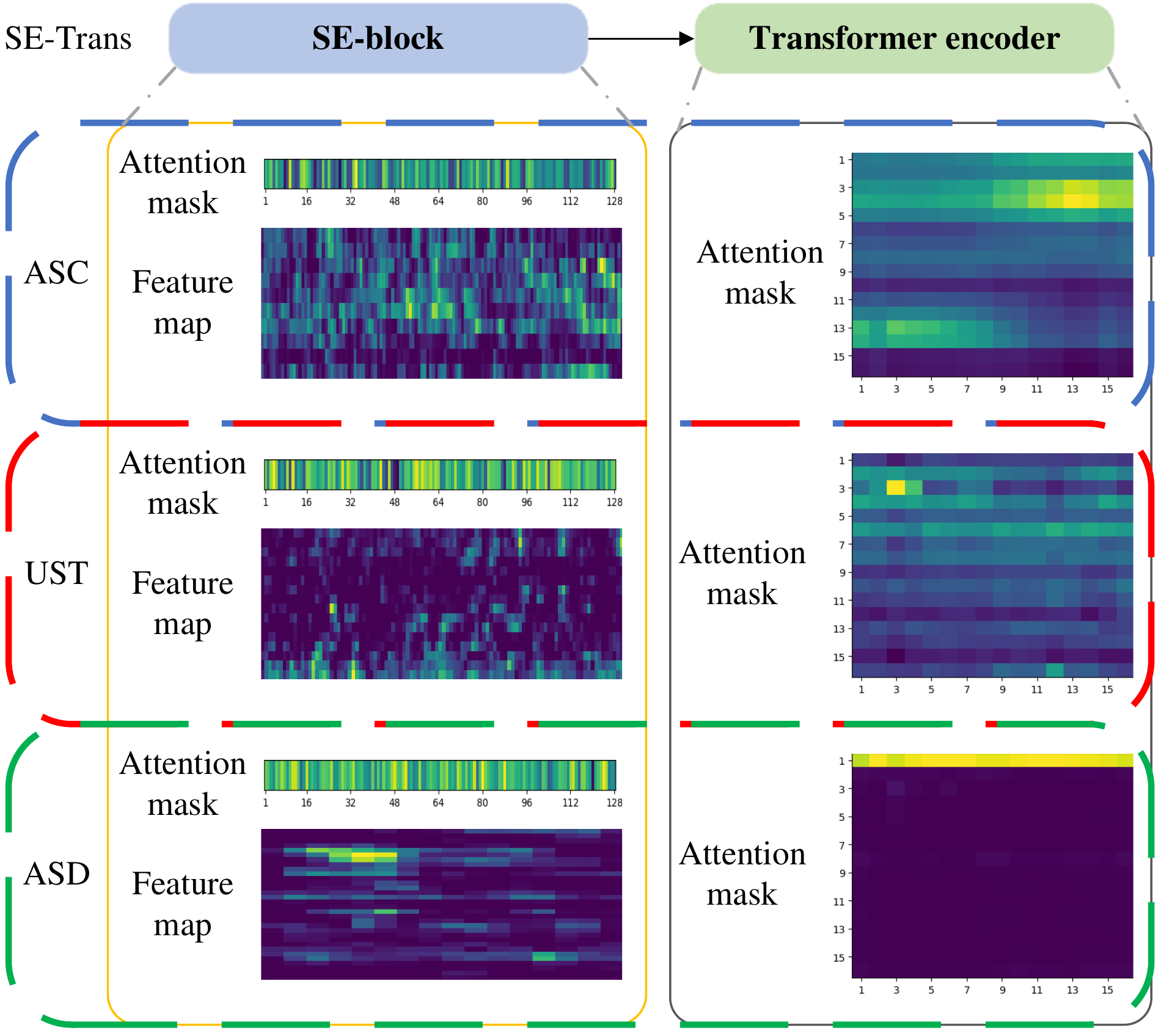}
\caption{Visualization of attention masks on ASC, UST and ASD.}\label{fig:Visual}
\end{figure} 

To study the effectiveness of the attention mechanism in the proposed SE-Trans, we further visualize the attention masks of the SE and Transformer encoder modules on ASC, UST, and ASD.
The attention mask of SE is the channel-wise weighted vector $\mathbf{w}$ in Eq. \ref{eq:w} and Fig. \ref{fig:SE block}, while the attention mask of the Transformer encoder is the mean attention of all the heads ($head_{i}$) in MHSA.
Moreover, we show the most important feature map  
from all of the feature maps in the second SE-block, i.e., the maximum value of the channel-wise weighted vector $\mathbf{w}$. 
The attention masks and feature maps are shown
in Fig. \ref{fig:Visual}.

In Fig. \ref{fig:Visual}, the attention masks of SE modules on different tasks are not the same. Specifically, the attention mask and feature map of ASC show that the feature map with global patterns is more important. 
For UST, the feature map shows more local but less global patterns.
While for ASD, attention is paid more to local patterns.
It is indicated that the global environmental background is meaningful for ASC task, the combination of different sounds is critical for UST, and the individual acoustic pattern is important for ASD.

For the attention mask of Transformer encoder, we can see that the relationship between temporal frames is different in the three tasks.
The attention mask of ASC tends to model the temporal dependencies on most of the frames. 
While the attention mask of UST pays more attention to the dependencies on some of the frames.
As for the attention mask of ASD, the attention is focused on a certain frame.

The visualization of attention masks illustrates distinct patterns for ASC, UST, and ASD tasks.
It can be interpreted that the introduced attention mechanisms of SE and Transformer are effective to improve the performance of cross-task models for ESR. 

\section{Conclusion}
\label{sec:conclusion}
This paper proposes a cross-task model to generally model acoustic knowledge across three different tasks of ESR.
In the model, an architecture based on two types of attention mechanisms is presented, namely SE-Trans.
This architecture exploits SE and Transformer encoder modules to learn the channel-wise importance and long sequence dependencies of the acoustic features.
We also adopt FMix to augment training data and extract robust sound representations efficiently.
Experiments show that our proposed cross-task model achieves state-of-the-art performance for ASC, UST, and ASD with lower computational resource demand.
Further analysis and visualization experiments explore the generality and individuality of acoustic modeling for ESR, and illustrate the effectiveness and robustness of the proposed model.

%\section*{ACKNOWLEDGMENT}


\ifCLASSOPTIONcaptionsoff
\newpage
\fi



\bibliographystyle{IEEEtran}
\bibliography{refers}
%\begin{thebibliography}{1}

\begin{IEEEbiography}[{\includegraphics[width=1in,height=1.25in,clip,keepaspectratio]{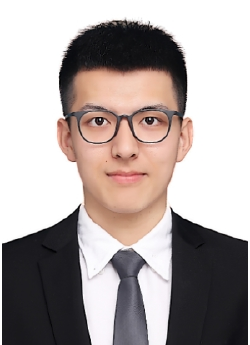}}]{Jisheng Bai}
received the B.S. degree in Detection Guidance and Control Technology from North University of China in 2017. 
He received the M.S. degree in Electronics and Communications Engineering from Northwestern Polytechnical University in 2020, where he is pursuing the Ph.D. degree. 
His research interests are focused on deep learning and environmental signal processing.
\end{IEEEbiography}

\begin{IEEEbiography}[{\includegraphics[width=1in,height=1.25in,clip,keepaspectratio]{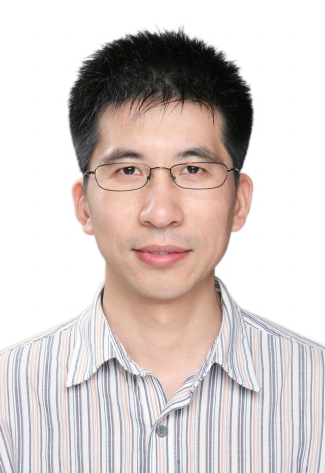}}]{Jianfeng Chen}
received the Ph.D. degree in 1999, from Northwestern Polytechnical University. 
From 1999 to 2001, he was a Research Fellow in School of EEE, Nanyang Technological University, Singapore.
During 2001-2003, he was with Center for Signal Processing, NSTB, Singapore, as a Research Scientist. 
From 2003 to 2007, he was a Research Scientist in Institute for Infocomm Research, Singapore. 
Since 2007, he works in College of Marine as a professor in Northwestern Polytechnical University, Xi’an, China. 
His main research interests include autonomous underwater vehicle design and application, array processing, acoustic signal processing, target detection and localization.
\end{IEEEbiography}

\begin{IEEEbiography}[{\includegraphics[width=1in,height=1.25in,clip,keepaspectratio]{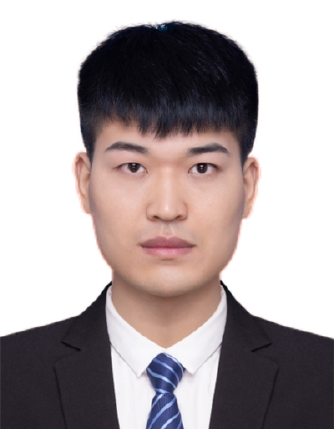}}]{Mou Wang}
(Student Member, IEEE) received the B.S. degree in electronics and information engineering from Northwestern Polytechnical University, China, in 2016, where he is pursuing the Ph.D. degree in information and communication engineering. His research interests include machine learning and speech signal processing. He was awarded Outstanding Reviewer of IEEE Transactions on Multimedia.
\end{IEEEbiography}

\begin{IEEEbiography}[{\includegraphics[width=1in,height=1.25in,clip,keepaspectratio]{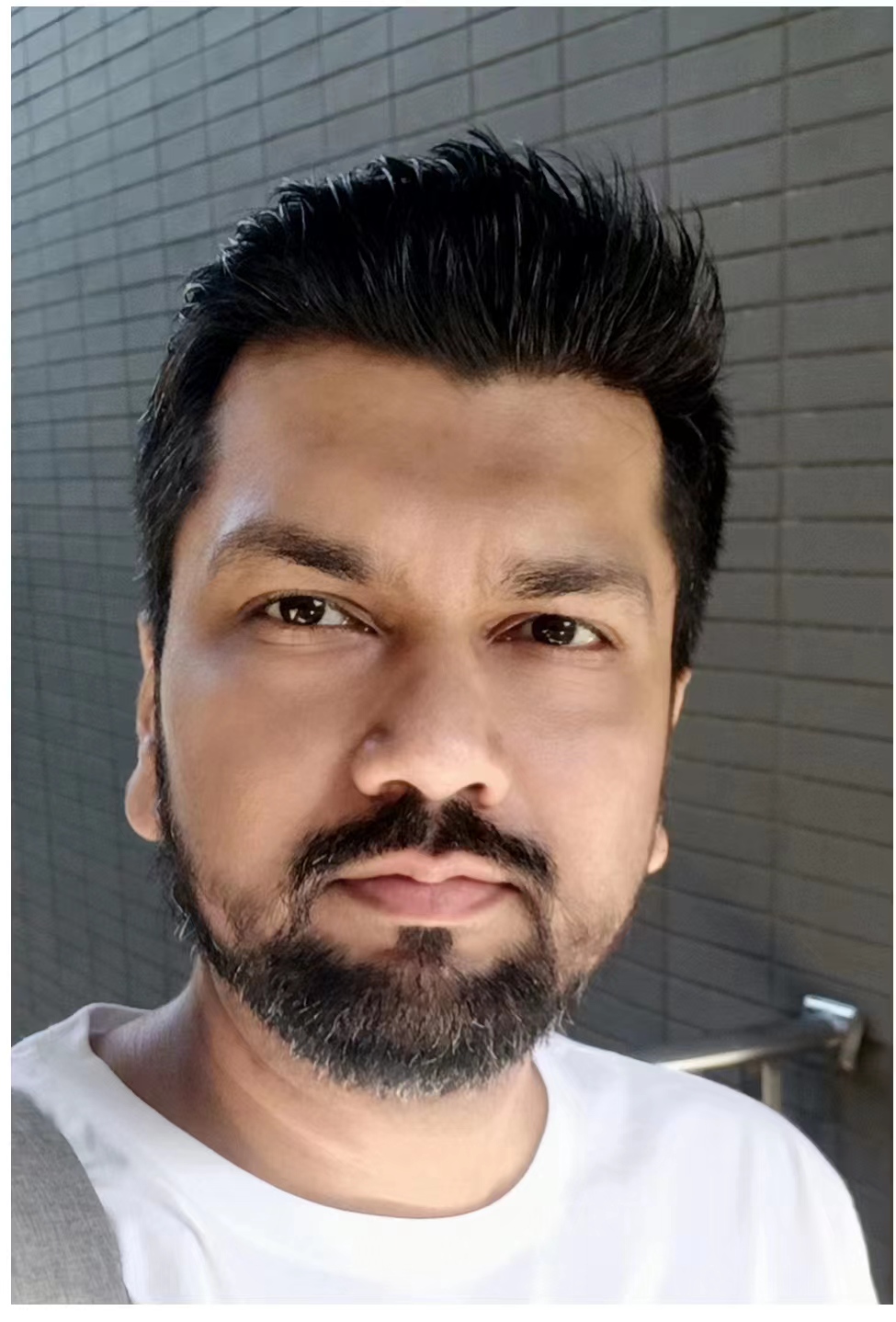}}]{Muhammad Saad Ayub}
received his B.E and M.Sc. degrees in 2010 and 2015 from National Univ. of Science and Technology Pakistan. He is currently working towards his PhD degree in School of Marine 
Science and Technology, Northwestern Polytechnical University, Xian. His research interests include 
acoustic signal analysis, multiple target localization, distributed acoustic networks, deep neural networks 
and formal analysis.
\end{IEEEbiography}

\begin{IEEEbiography}[{\includegraphics[width=1in,height=1.25in,clip,keepaspectratio]{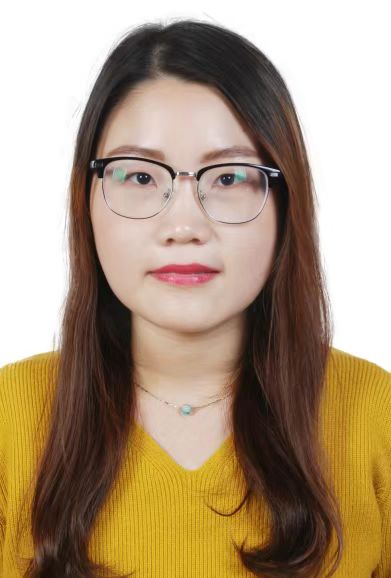}}]{Qingli Yan}
received her Master’s degree and Ph.D. degree in signal processing from Northwestern Polytechnical University in 2015 and 2019. Since 2019, she has been a lecturer in the School of Computer Science \&
Technology, Xi’an University of Posts \& Telecommunications. Her research interests include wireless sensor network, target localization and tracking.
\end{IEEEbiography}

\end{document}